\documentclass[fleqn,usenatbib]{mnras}

\usepackage{newtxtext,newtxmath}

\usepackage[T1]{fontenc}
\usepackage{amsmath}

\DeclareRobustCommand{\VAN}[3]{#2}
\let\VANthebibliography\thebibliography
\def\thebibliography{\DeclareRobustCommand{\VAN}[3]{##3}\VANthebibliography}

\usepackage{graphicx}	
\usepackage{amsmath}	
\usepackage{threeparttable}

\newcommand{\pchance}{$P_{\rm chance}$}
\newcommand{\msol}{$M_{\odot}$~}

\newcommand{\sfrunits}{\msol yr$^{-1}$}
\newcommand{\alphalotss}{$\alpha_{\rm LoTSS}$}
\newcommand{\alphawide}{$\alpha_{\rm wide}$}
\newcommand{\qir}{$q_{\rm IR}$}

\title[Finding LoTSS of hosts for GRBs]{Finding LoTSS of hosts for GRBs: a search for galaxy - gamma-ray burst coincidences at low frequencies with LOFAR}

\author[R. A. J. Eyles-Ferris and R. L. C. Starling]{
R. A. J. Eyles-Ferris$^{1}$\thanks{E-mail: raje1@leicester.ac.uk}
and R. L. C. Starling$^{1}$
\\
$^{1}$School of Physics and Astronomy, University of Leicester, University Road, Leicester, LE1 7RH, UK\\
}

\date{Accepted XXX. Received YYY; in original form ZZZ}

\pubyear{2023}

\begin{document}
\label{firstpage}
\pagerange{\pageref{firstpage}--\pageref{lastpage}}
\maketitle

\begin{abstract}
The LOFAR Two-Metre Sky Survey (LoTSS) is an invaluable new tool for investigating the properties of sources at low frequencies and has helped to open up the study of galaxy populations in this regime. In this work, we perform a search for host galaxies of gamma-ray bursts (GRBs). We use the relative density of sources in Data Release 2 of LoTSS to define the probability of a chance alignment, \pchance, and find 18 sources corresponding to 17 GRBs which meet a \pchance<1\% criterion. We examine the nature and properties of these radio sources using both LOFAR data and broadband information, including their radio spectral index, star formation rate estimates and any contributions from active galactic nucleus emission. Assuming the radio emission is dominated by star formation, we find that our sources show high star formation rates ($10^1$--$10^3$\sfrunits) compared with both a field galaxy sample and a sample of core-collapse supernova hosts, and the majority of putative hosts are consistent with ultraluminous infrared galaxy (ULIRG) classifications. As a result of our analyses, we define a final sample of eight likely GRB host candidates in the LoTSS DR2 survey.
\end{abstract}




\section{Introduction}

The properties of gamma-ray bursts (GRBs) are the result of their progenitors and therefore the environment in which they have evolved. The majority of GRBs discovered to date are long GRBs, bursts in which 90\% of the isotropic equivalent energy is detected over a period longer than two seconds. These GRBs are driven by the collapse of massive stars and hence are strongly connected to star formation. As such, while GRBs are cosmological and are found across redshifts up to 9.4 \citep{Cucchiara11}, the peak of their redshift distribution is also commensurate with the peak of star formation \citep[$z\sim 2.2$,][]{Evans09,Fynbo09,Jakobsson12}. However, despite this strong connection, long GRBs are not unbiased tracers of star formation \citep{Levesque14}. Long GRBs have been found to be hosted in a wide variety of galaxy types but there is evidence of a bias towards galaxies with lower masses, high specific star formation and low metallicities \citep[e.g.][]{Christensen04,Savaglio09,Lyman17}. These environments are ideal for forming collapsars, the progenitors of long GRBs, and other studies have suggested that higher metallicities actually suppress collapsar formation \citep[e.g.][]{Yoon06,Woosley06,Perley16}. Short GRBs, where 90\% of the isotropic equivalent energy is detected over a period shorter than two seconds, have different progenitor systems and are driven by compact binary mergers. They therefore display a greater variety in their host galaxy properties and do not require active star formation. There is also a minority of short GRBs that appear to lack hosts - most likely due to natal kicks ejecting them from their original hosts and resulting in offsets of tens of kiloparsecs \citep[e.g.][]{Tunnicliffe14,Mandhai22,Fong22}.

It is therefore clear that a galaxy's star formation rate (SFR) has a significant impact on its likelihood to host long GRBs in particular. There are several methods to measure SFR including at UV and optical wavelengths. However, these methods are significantly affected by dust extinction in the target galaxy. This extinction, and its effects on SFR measurements, is often hard to constrain and leads to significant biases. Star formation also drives radio emission, however, which is significantly less affected by dust extinction. This therefore offers an opportunity for dust unbiased measurements of long GRB hosts.

There are factors to consider when using radio data in this way, however. While radio emission is a good tracker of star formation, a significant proportion of extragalactic radio sources are active galactic nuclei (AGN). AGN can also be a significant contaminant in sources that are also highly starforming. It is therefore important to minimise AGN contamination to accurately measure the properties of any galaxies identified. There is also the possibility of lingering emission from any GRB afterglows further contaminating the sample. This has occurred in other studies \citep[e.g.][]{Stanway14} although we do not expect it to be a contributor in here due to the low frequency of the regime probed and the historic nature of the GRBs investigated.

We also need to consider other biases inherent to such a survey. The luminosity of the radio emission from star formation is strongly dependent on the SFR. In a blind search for such sources, therefore, we may expect to only detect galaxies with very elevated SFRs. While we expect long GRB hosts to have high SFRs, it could be difficult to determine whether the hosts detected via radio are representative of the entire population or comprise an outlying subsample of extreme starforming galaxies. Alternatively, a large proportion of long GRB hosts with lower, albeit still high, SFRs could remain undetected and the search would be inherently incomplete. Due to star formation no longer being necessary, compared to long GRB hosts, short GRB hosts are more likely to be radio-quiet despite their lower mean redshift \citep[$z\lesssim1$,][]{Klose19,Fong22}. This means a radio search is also unlikely to detect them and lead to greater incompleteness across the broader GRB population.

Radio searches for GRB hosts have proven to be productive in past studies, however \citep[e.g.][]{Berger03,Michalowski12,Stanway14,Li15,Perley15,Perley16,Klose19}. At GHz frequencies, these galaxies typically have measured SFRs much higher than those obtained through UV/optical methods \citep[e.g.][]{Christensen04,Michalowski12}. Long GRBs are also commonly hosted in powerful luminous infrared galaxies (LIRGs) and ultra-luminous infrared galaxies (ULIRGs) and the highly dusty nature of these galaxies means radio is the only way to recover obscured star formation. In these cases, SFRs of hundreds to over a thousand \sfrunits have been recovered \citep{Perley17,Hsiao20}. The fraction of GRB hosts with such dust obscuration is still unclear \citep[e.g.][]{Gatkine20} and radio surveys offer a way to constrain this further. While some of the differences between UV/optical SFRs and radio SFRs have been attributed to afterglow or AGN contamination, it is clear radio is a valuable tool to investigate these galaxies.

A new window into the low frequency behaviour of galaxies has recently opened with the second Data Release of the LOFAR Two-metre Sky Survey \citep[LoTSS DR2,][]{Shimwell22}. LoTSS DR2 covers 6335 square degrees in two regions centred on 12h45m00s +44\degr30'00" (RA-13 region) and 1h00m00s +28\degr00'00" (RA-1 region), approximately 27\% of the northern hemisphere. The frequency range is significantly lower than most radio surveys at 120--168 MHz subdivided into three 16 MHz wide bands, a regime almost entirely unprobed in studies of GRB host galaxies. The survey achieves a resolution of 6" and RMS limits of 74 $\mu$Jy and 106 $\mu$Jy in the RA-13 and RA-1 regions respectively. This has allowed 4.4 million sources to be detected, and in this paper, we investigate these catalogues to identify associations between LoTSS sources and GRBs detected with the X-ray Telescope \citep[XRT,][]{Burrows05} on board the {\it Neil Gehrels Swift Observatory} \citep[{\it Swift},][]{Gehrels04}. These GRBs are extremely well localised to within $\le$ 2 arcsec 90\% of the time \citep{Evans09}, ideal for crossmatching to LoTSS DR2.

In Section \ref{sec:assoc}, we present our crossmatching method and initial associations. We further discuss our method in Section \ref{sec:selectcomplete}, focussing on any selection biases and its completeness. The properties of our matches are investigated and presented in Section \ref{sec:host_properties} and these are combined with our discussion of our methodology to produce a final sample of GRB associations in \ref{sec:final_sample}. Finally, we summarise in Section \ref{sec:conc}.

Throughout this paper, we give errors to 1-$\sigma$ and adopt a flat $\Lambda$CDM cosmology with $H_0 = 71$ km\,s$^{-1}$\,Mpc$^{-1}$, $\Omega_m = 0.27$ and $\Omega_\Lambda = 0.73$.

\section{Associating GRBs and possible hosts}
\label{sec:assoc}

\subsection{Crossmatching}
\label{sec:crossmatch_Tunnicliffe}

\begin{figure*}
    \centering
    \includegraphics[width=\textwidth]{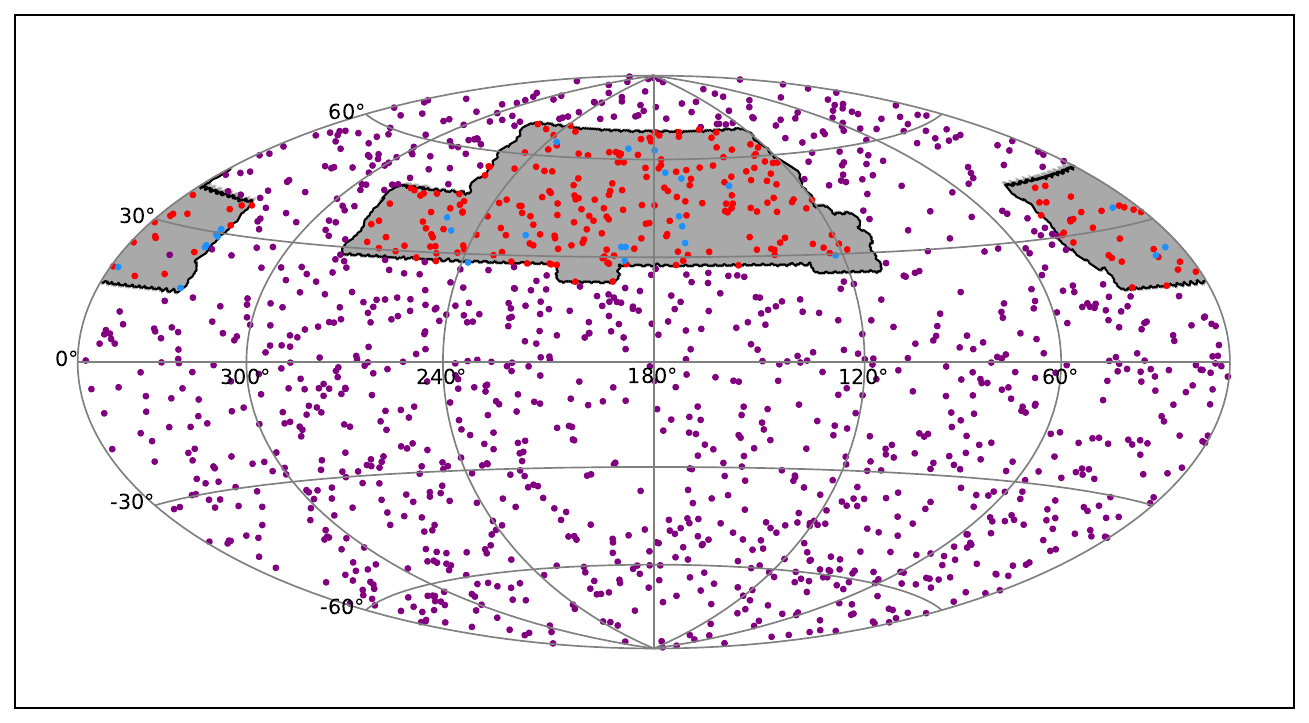}
    \caption{The coverage of LoTSS DR2 shown in grey. The image is focussed on the RA-13 region and the RA-1 region is split. Red point and blue points indicate long and short GRBs respectively within the LoTSS DR2 footprint while purple points indicate GRBs of both types outside the footprint.}
    \label{fig:moc}
\end{figure*}

Our sample of GRBs was taken from the live \textit{Swift}-XRT GRB Catalogue\footnote{\url{https://www.swift.ac.uk/xrt_live_cat/}} \citep{Evans09} hosted by the UK \textit{Swift} Science Data Centre (UKSSDC) up to 15 July 2022. This consisted of 1489 GRBs and we found 280 of them (253 long GRBs and 27 short GRBs) were located within the LoTSS DR2 footprint, defining the footprint as the area within one degree of at least one LoTSS DR2 source, as shown in Figure \ref{fig:moc}. We note that two sources apparently within the LoTSS footprint are excluded by this criterion but this is most likely due to blanked regions within the footprint\footnote{see \url{https://lofar-surveys.org/dr2_release.html}}. We matched this sample to the \texttt{lotss\_dr2.main\_sources} table \citep{lotss_dr2}, selecting all LoTSS sources within 10 times the 90\% XRT error region for each GRB. We then calculated the probability of chance alignment, \pchance, following the method of \citet{Tunnicliffe14}. In that work, GRBs were matched to possible host galaxies in the Sloan Digital Sky Survey (SDSS) by examining the correlation of flux and source density in a sample of the SDSS. The \pchance~of possible matches was then evaluated by comparison with this overall population.

We generated 15,000 random points within the footprint of LoTSS DR2 and measured the separations from these points to the nearest sources above various total flux thresholds. We classified these sources by cross-matching to the NASA/IPAC Extragalactic Database\footnote{\url{https://ned.ipac.caltech.edu/}} (NED) and the Set of Identifications, Measurements and Bibliography for Astronomical Data \citep[SIMBAD,][]{Wenger00} database\footnote{\url{https://simbad.u-strasbg.fr/simbad/}}, assuming the LoTSS DR2 source to correspond to the closest object within 20". These cross matches were then used to select only galaxies from the sample and we found that the relationship between the separations from the random points, $\delta x$, and the flux of the galaxies, $F$, was reasonably well fit as
\begin{equation}
    \log(\delta x) = \alpha \log(F) + c
\end{equation}
where $\alpha = 0.417\pm0.002$ and $c = 2.153\pm0.005$. We also took a comparison sample of field galaxies from this population, selecting sources with SIMBAD galaxy counterparts with sufficiently low separations for their \pchance<1\%. We further subdivided this into active and inactive galaxies using their SIMBAD classifications. We refer to these samples later in this paper as \textsc{ActiveField} and \textsc{InactiveField}, respectively.

\begin{figure*}
    \centering
    \includegraphics[width=\textwidth]{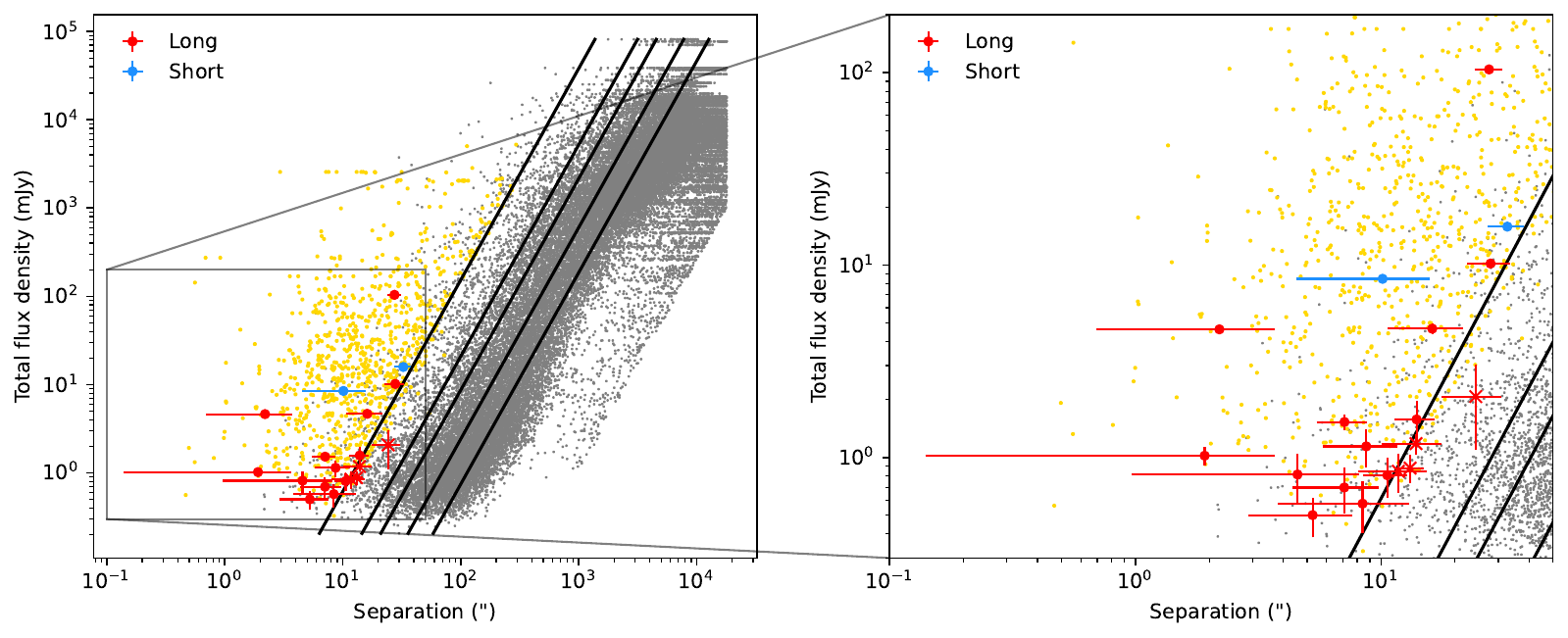}
    \caption{The correlation between separation and flux density for galaxies in the LoTSS DR2 release. Grey indicates matches to 15,000 random points and the black lines indicate the 1st, 5th, 10th, 25th and 50th percentile contours. Matches to our CCSNe sample are plotted in gold and matches to long and short GRBs are plotted in red and blue respectively. For the GRB matches, circles indicate sources where \pchance~is less than 1\% and crosses indicate sources where \pchance~is consistent within errors of being less than 1\%. The right hand panel zooms in to more clearly show our matches.}
    \label{fig:crossmatch}
\end{figure*}

We used this fit to define the percentile contours of the distribution and we plot the 1st, 5th, 10th, 25th and 50th percentiles in Figure \ref{fig:crossmatch}. For each source matched to a GRB, \pchance~was therefore determined by which percentile contour the matched source lay on. We calculated these for all of our matched sources and selected a threshold of \pchance<1\%. We found that twelve long GRBs and one short GRB had at least one match meeting this threshold, while a further four long GRBs had matches consistent with the threshold when accounting for errors. We plot the flux density and separation from the GRBs of our crossmatched sources in Figure \ref{fig:crossmatch} and summarise them in Table \ref{tab:crossmatch}. We also extracted images of each LoTSS source and the position of their potential match from the full LoTSS mosaics and present them in Figure \ref{fig:mosaics}. All of the LoTSS sources are most likely galaxies, with the majority having known counterparts in SIMBAD. Two GRBs, 050509B and 081025, were found to have multiple matches. We discuss these sources further below. The 17 matched GRBs represent 6.0\% of the 280 GRBs within the footprint. For the two GRB types, this breaks down into 16/253 or 6.3\% of long GRBs and 1/27 or 3.8\% of short GRBs.

This method can lead to crossmatches at large angular separations, up to $\sim35$". This is primarily due to the lower source density at higher flux density limits and therefore greater separations are allowed. Such large separations are possible. For instance, the formation of the compact binary progenitors of short GRBs induces a significant kick velocity such that short GRBs can occur a significant distance from their original hosts. For both short and long GRBs, there are also the possibilities that the GRBs are occurring in the outskirts of extended galaxies or that they are located at low redshifts where a large angular separation does not translate to a large physical separation. This latter case would also account for the high brightness of some putative hosts but is also inconsistent with the observed gamma-ray fluence of several of the bursts. Finally, there is also the possibility of the positional accuracy playing a significant role.

Alternatively, many of the LoTSS sources are extended, consistent with their galactic nature, and as such the separations may also be dependent on the angular size of the host. We calculated the extent of each LoTSS source towards the centre of the associated GRB's XRT error region assuming the source to be Gaussian in nature and include this in Table \ref{tab:crossmatch}. Note that we used the major and minor axes of the sources after deconvolution with the beam, while the images in Figure \ref{fig:mosaics} are derived from the raw mosaics.  We also derived the normalised separation, i.e. the separation divided by the extent. This has previously been investigated at optical wavelengths by \citet{Blanchard16}, who found that for long GRBs, the normalised separation was typically $\lesssim$one but could range up to $\sim$nine in extreme cases. Our normalised separations are generally somewhat larger than this but are not incompatible within errors. However, there are a number of sources with significantly larger separations, further reducing the likelihood of these being accurate crossmatches.

\subsubsection{GRBs with multiple matches}

For our only short GRB, GRB 050509B, two separate LoTSS sources were consistent with being matches according to our \pchance~criterion. GRB 050509B is a relatively well studied short GRB and it is likely to have occurred within a cluster \citep{Barthelmy05,Pedersen05}. Multiple matches are therefore not surprising. While the afterglow was only detected in X-rays, the GRB was sufficiently localised for Very Large Telescope observations to confirm the likely host is 2MASX J12361286+285858026 \citep{Hjorth05a,Hjorth05,CastroTirado05a}, a large elliptical galaxy with a UV-measured SFR of $<0.2$ \sfrunits~\citep{Gehrels05}. As shown in Figure \ref{fig:050509b}, one of our matched sources, ILT J123613.00+285902.9, is spatially consistent with this galaxy. The other LoTSS source, ILT J123612.26+285929.2, is likely unrelated to GRB 050509B. However, we still evaluate its properties as it could be part of the same cluster as 2MASX J12361286+285858026 and therefore coevolved with it.

Upon inspection of the LoTSS mosaics, we also found that the counterpart to GRB 081025, ILT J162131.13+602837.1, appeared to be a double source, as shown in Figure \ref{fig:mosaics}. From archival 2MASS images, we determined that it was two distinct galaxies with insufficient angular separation to be resolved as individual sources in the LoTSS pipeline. However, the pipeline also outputs the separate Gaussians that make up each source and we were able to identify the two Gaussians, here designated A and B, matching each component of ILT J162131.13+602837.1, both of which individually met the \pchance~threshold. We therefore treated these as separate sources in our analysis.

\subsubsection{Redshifts and physical scales}
\label{sec:crossmatch_z}

The redshifts of our sample are important for both accurately measuring the properties of the LoTSS sources but for our crossmatching process. Four of the matched GRBs already had measured redshifts, two each from spectroscopy of their afterglows and of their probable hosts. We identified further redshifts from catalogue crossmatches comprising two from SIMBAD and five photometric redshifts from SDSS. Of our 18 matches, therefore, we identify 12 possible redshifts.

Based on these redshifts, we calculated both the physical separation of the GRB from the LoTSS source and the apparent physical extent of the sources. These values are also given in Table \ref{tab:crossmatch}.

The separations are generally significantly higher than the typical results of a $\sim$few kpc identified previously for long GRBs \citep[e.g.][]{Bloom02,Blanchard16,Lyman17}. However, we note the previous results are derived using optical data, and some differences might therefore be expected, and that our separations typically have relatively large errors. Nevertheless, the more extreme values are likely indicative of inaccuracies in the crossmatching process.

On the other hand, the behaviour of short GRBs differs and a more significant separation would often be expected. This is due to the nature of short GRBs, the progenitors of which can be imparted a significant kick velocity through their formation mechanism, and therefore have large separations from their original host. Observationally, these separations are a few kpc greater than those of long GRBs \citep[e.g.][]{Tunnicliffe14,Fong22}. From simulations, the distribution of separations could be expected to peak at around 10 kpc and can reach to over 100 kpc \citep{Mandhai22}. Our matches to GRB 050509B are both consistent with this behaviour.

\subsection{Core-collapse supernovae sample}

To provide an additional comparison sample to our putative GRB hosts, we also derived a sample of core-collapse supernovae (CCSNe) host candidates following the same procedure. CCSNe are a well studied population of which long GRBs could represent a subsample.

CCSNe are generally found at significantly lower redshifts than long GRBs and typically belong to types Ib, Ic or II. In previous studies, there have been found to be significant differences between hosts of CCSNe and long GRBs \citep[e.g.][and references therein]{Levesque14}, however, these differences have been found to vary between types. For instance, long GRBs tend to be located within UV bright regions of their hosts, consistent with high degrees of star formation \citep[e.g.][]{Svensson10}. The resultant offset distribution is more similar to Ib or Ic CCSNe than type II \citep{Blanchard16,Lyman17}. Similarly, Ic and long GRBs both trace optical \textit{g}-band emission within their hosts with other CCSNe displaying weaker associations \citep{Kelly08}. For all CCSNe types, the luminosity and morphology distribution of their hosts is also substantially different than for long GRBs. \citet{Fruchter06} found that CCSNe hosts tend to be both more luminous and more regular than long GRB hostss. This supported by \citet{Svensson10}'s conclusion that CCSN hosts tend towards massive spirals. Compared to the low metallicity nature of long GRB hosts, CCSN hosts are also typically much more metal rich \citep[e.g.][]{Levesque10}.

We took the sample of the sadly defunct Open Supernova catalogue\footnote{\url{https://github.com/astrocatalogs/supernovae}} which includes events up to April 8 2022, and selected all supernovae of suggested type Ib, Ic and II. This resulted in 1162 CCSNe within the LoTSS DR2 footprint, which we defined as above.  We crossmatched these objects to LoTSS DR2 with a \pchance~threshold of 1\% which resulted in 877 matches corresponding to 664 SNe, or 57.1\% of the sample within the LoTSS DR2 footprint.

We found the low redshifts of the CCSNe population to have a significant effect on the crossmatching process. Such nearby sources have correspondingly high flux densities and the angular offsets of the supernovae from the LoTSS sources were found to be significantly greater than those of the GRB sample, in some cases $>100$". However, these angular offsets still correspond to physical offsets of $\sim$tens of kpc or less. The nine times larger proportion of CCSNe matches compared to GRB matches is therefore not surprising.

The proximity of the CCSNe sample is also likely responsible for the large number of supernovae with multiple LoTSS matches as many supernovae are close enough for their hosts' structure to be resolved. Such galaxies will be detected as multiple sources in the LoTSS pipeline as the islands of flux are sufficiently separated.

\section{Selection effects and completeness}
\label{sec:selectcomplete}

\subsection{Selection biases}
\label{sec:selection}

It is important to consider the biases inherent in a radio selection of host galaxies. As most long GRBs are associated with star-forming galaxies, it might be expected that radio emission from their hosts is dominated by the synchrotron radiation associated with star formation but that the minimum SFR probed will depend on the survey depth and galaxy redshift. \citet{Gurkan18} found the SFR, $\psi$, and luminosity at 150 MHz, $L_{150}$, to be correlated as
\begin{equation}
\label{eq:SFR_Gurkan}
    L_{150} = L_1 \psi_{\rm G}^{\beta}
\end{equation}
where $L_1 = 10 ^ {22.06 \pm 0.01}$ W Hz$^{-1}$ is the luminosity at 150 MHz of a source with an SFR of 1 \sfrunits~and $\beta=1.07 \pm 0.01$. \citet{Smith21} also derive SFRs for 150 MHz emission, finding a similar prescription:
\begin{equation}
\label{eq:SFR_Smith}
\begin{aligned}
    \log L_{150} = & (0.90 \pm 0.01) \log (\psi_{\rm S} / M_{\odot}~{\rm yr}^{-1}) + \\ & (0.33 \pm 0.04)
   \log (M_{\rm gal} / 10 ^{10} M_{\odot}) + 22.22 \pm 0.02
\end{aligned}
\end{equation}
for $L_{150}$ in W Hz$^{-1}$. We note, however, that this method requires the mass of the galaxy.

\begin{figure}
    \centering
    \includegraphics[width=\columnwidth]{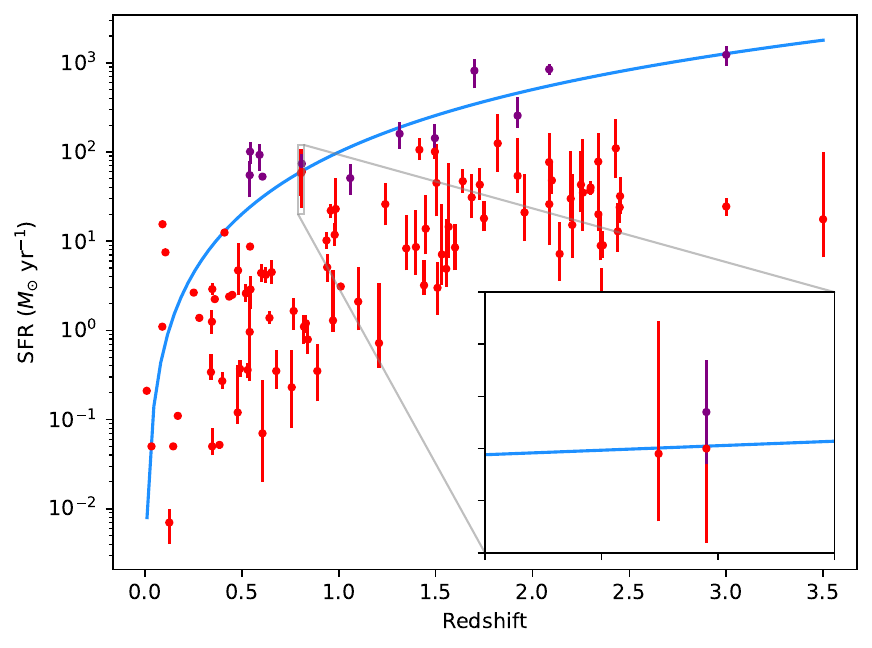}
    \caption{The required SFR to reach 0.30 mJy at 150 MHz for a given redshift (blue curve). The UV/optical derived SFRs for a sample taken from \citet{Wang20} and the live \textit{Swift}-XRT GRB Catalogue \citep{Evans09} are overlaid in red. Radio derived SFRs from \citet{Perley13}, \citet{Stanway14} and \citet{Perley15} are overlaid in purple. The inset zooms into GRBs 051022 and 100816A as they are coincidentally close and cannot be distinguished in the full plot.}
    \label{fig:sfr_curve}
\end{figure}

Equation \ref{eq:SFR_Gurkan} can therefore be used to evaluate these selection effects. We used the grey field sources in Figure \ref{fig:crossmatch} to derive an approximate detection limit for LoTSS DR2. We found that only 0.1\% of sources have total flux densities <0.30 mJy. We therefore take this to be an appropriate and relatively conservative assumption for the detection limit. Using Equation \ref{eq:SFR_Gurkan}, we derive the SFR required to reach this flux density up to $z=3.5$ and plot it in Figure \ref{fig:sfr_curve}.

We also compare this function to a sample of long GRB hosts with SFRs estimated through optical or UV methods. Such methods are significantly more impacted by the effects of dust obscuration than radio methods and may therefore result in underestimated SFRs. We would therefore expect radio methods to favour both higher SFRs and lower redshfits. The sample is taken from GRBs in both the catalogue of \citet{Wang20} and the live \textit{Swift}-XRT GRB Catalogue \citep{Evans09}. This yields 92 GRB hosts with no crossover with our crossmatched sample, plotted in red in Figure \ref{fig:sfr_curve}. We also plot 12 long GRB hosts with radio measured SFRs taken from \citet{Perley13}, \citet{Stanway14} and \citet{Perley15}, noting that GRB 051022 is present in both samples with measured SFRs consistent within errors. This comparison demonstrates that radio detections are intrinsically biased towards hosts with atypically high SFRs compared to the UV/optically measured sample. The redshift distribution of sources we might expect to detect also differs from the full population with a clear bias towards lower redshifts, probably due to the SFRs required being significantly less extreme than for most of the sample. We do note, however, that 6 of the 92 UV/optically derived SFRs do reach the threshold. The radio emission of some putative hosts may also be enhanced by components other than star formation. The most obvious source of such contamination is likely to be AGN and we investigate this in Section \ref{sec:agn}.

\subsection{Completeness}
\label{sec:completeness}

The sample of GRB hosts from \citet{Wang20} also allow the completeness of our methods to be evaluated. Of the 92 GRBs in the UV/optical sample, six ($\sim$6.5\%) GRB hosts do appear to reach the threshold for detection. We also found that two hosts, those of GRBs 051022 and GRB 100816A, lie within the LoTSS DR2 footprint and have UV/optically derived SFRs of $60.0^{+12.0}_{-36.2}$ \sfrunits~\citep{CastroTirado07,Svensson10,Kruhler15} and $58.0^{+51.0}_{-26.0}$ \sfrunits~\citep{Kruhler15} respectively. GRB 051022 has also been observed at 5.227 GHz which yielded a slightly enhanced SFR of $74.0^{+20.0}_{-20.0}$ \sfrunits~\citep{Perley13}. We therefore predict flux densities at 144 MHz using Equation \ref{eq:SFR_Gurkan} of $0.30^{+0.18}_{-0.19}$ mJy for GRB 051022's counterpart and $0.29^{+0.28}_{-0.14}$ mJy for GRB 100816A's. These values are consistent with being below our estimated detection limit and no such counterparts were identified in the LoTSS DR2 catalogues for either GRB.

To confirm these non-detections, we examined the mosaics at the positions of these GRBs and for GRB 100816A, found no evidence of a source. In contrast, we did identify a bright source approximately 16" away from GRB 051022, ILT J235603.13+193633.0, as shown in Figure \ref{fig:051022}. This source was not previously selected by our criteria, due to the well localised nature of GRB 051022 and its XRT error region of 1.5". We note, however, that its \pchance~would meet the 1\% threshold.

We therefore examined the GRB Coordinates Network Circulars for GRB 051022 and found \citet{Cameron05} reported two radio sources, a point source coincident with the XRT error region and a previously reported optical source \citep{CastroTirado05a}, and a large extended source to the North West of the error region. Further observations indicated the point source to indeed be the GRB afterglow \citep{vanderHorst05,deUgartePostigo05,Rol07}. ILT J235603.13+193633.0 is, however, consistent with the extended source and we therefore conclude it is unrelated to GRB 051022.

While we cannot directly measure the SFR of these two hosts, our non-detections do constrain their behaviour somewhat. In particular, the galaxies' radio SFRs cannot be significantly greater than those measured through UV/optical methods. This is consistent with the work of \citet{Greiner16}, who observed a sample of 11 long GRB hosts at 2.1 GHz and 3 GHz. No hosts were detected and the limits were sufficient to establish that any radio detectable star formation would be only be a factor of two to three greater than that measured by UV/optical methods.  While the redshifts of the two non-detected sources are only constrained to $z\sim0.8$, our results agree with \citeauthor{Greiner16}'s conclusion that only a small fraction of the star formation in these galaxies may be obscured by dust.

The non-detections provide weak constraints on the completeness of our method. Although subject to small number statistics, they suggest that the fraction of long GRB host candidates we might identify in LoTSS DR2 is $\lesssim$6.5\%, consistent with the 6.3\% we find crossmatches for. As discussed in Section \ref{sec:selection} and later, it is also plausible that our methods actually lead to spurious associations and further evaluation and elimination of some host candidates must be performed.

\begin{table*}
\centering
\caption{A summary of the GRBs and crossmatches we identify in Section \ref{sec:crossmatch_Tunnicliffe} ordered by GRB class and \pchance. GRBs are classified as short (S) and long (L) based on the standard observed $t_{90}=2$ s cutoff. Note that the extent columns refer to the LoTSS source's apparent extent from its centre towards the GRB position rather than the maximum size of the source.}
\begin{threeparttable}
\renewcommand{\arraystretch}{1.4}
\begin{tabular}{cccccccccc}
    \hline
    GRB & Class & LoTSS match & Redshift & Angular & Angular & Normalised & Physical & Physical & \pchance~\\
    & & & & separation (") & extent (") & separation & separation (kpc) & extent (kpc) & (\%) \\
    \hline
200716C & L & ILT J130402.61+293839.3 & $0.341^{1}$ & $2.20\pm1.50$ & $2.05^{+0.11}_{-0.15}$ & $1.1^{+0.9}_{-0.7}$ & $10.6\pm7.2$ & $9.87^{+0.55}_{-0.72}$ & $0.01^{+0.01}_{-0.01}$ \\
201229A & L & ILT J140245.38+481150.7 & --- & $1.91\pm1.77$ & $0.75^{+0.70}_{-0.52}$ & $2.5^{+13.5}_{-2.5}$ & --- & --- & $0.02^{+0.07}_{-0.02}$ \\
190211A & L & ILT J130635.93+415811.3 & --- & $27.49\pm3.51$ & $1.77^{+0.01}_{-0.01}$ & $15.5^{+2.1}_{-2.1}$ & --- & --- & $0.09^{+0.03}_{-0.02}$ \\
060123 & L & ILT J115847.46+453050.9 & $0.562^{2}$ & $4.57\pm3.60$ & $4.81^{+1.56}_{-1.94}$ & $1.0^{+1.9}_{-0.8}$ & $25.9\pm20.5$ & $27.28^{+8.87}_{-11.01}$ & $0.15^{+0.52}_{-0.15}$ \\
091130B & L & ILT J133235.11+340524.3 & $0.282^{3}$ & $7.10\pm1.63$ & $1.08^{+0.46}_{-0.19}$ & $6.6^{+3.2}_{-3.0}$ & $30.0\pm6.9$ & $4.58^{+1.95}_{-0.79}$ & $0.20^{+0.16}_{-0.09}$ \\
110903A & L & ILT J130815.16+585857.2 & $0.311^{3}$ & $5.26\pm2.38$ & $<0.36$ & $>8.0$ & $23.8\pm10.8$ & $<1.61$ & $0.31^{+0.58}_{-0.23}$ \\
081025 & L & ILT J162131.13+602837.1 A & --- & $16.15\pm5.49$ & $4.05^{+0.48}_{-0.42}$ & $4.0^{+2.0}_{-1.6}$ & --- & --- & $0.41^{+0.44}_{-0.25}$ \\
150213B & L & ILT J165348.36+341126.2 & $0.704^{3}$ & $7.08\pm2.73$ & $<0.74$ & $>5.9$ & $50.7\pm19.6$ & $<5.27$ & $0.41^{+0.70}_{-0.29}$ \\
220412A & L & ILT J081553.01+302035.9 & $0.402^{4}$ & $8.70\pm2.92$ & $2.14^{+1.44}_{-0.78}$ & $4.1^{+4.5}_{-2.5}$ & $46.6\pm15.7$ & $11.47^{+7.72}_{-4.16}$ & $0.42^{+0.58}_{-0.26}$ \\
081025 & L & ILT J162131.13+602837.1 B & --- & $27.89\pm5.47$ & $5.86^{+0.52}_{-0.50}$ & $4.8^{+1.5}_{-1.2}$ & --- & --- & $0.68^{+0.33}_{-0.27}$ \\
071020 & L & ILT J075839.09+325133.9 & $2.146^{5}$ & $8.41\pm4.60$ & $<1.05$ & $>3.6$ & $70.7\pm38.7$ & $<8.82$ & $0.74^{+1.57}_{-0.63}$ \\
110521A & L & ILT J080031.58+454950.2 & --- & $13.95\pm2.61$ & $1.27^{+1.10}_{-0.47}$ & $11.0^{+9.7}_{-6.2}$ & --- & --- & $0.91^{+0.63}_{-0.44}$ \\
060206 & L & ILT J133144.19+350305.6 & $4.050^{6}$ & $10.60\pm2.17$ & $3.02^{+0.74}_{-1.17}$ & $3.5^{+3.4}_{-1.3}$ & $74.7\pm15.3$ & $21.31^{+5.21}_{-8.23}$ & $0.92^{+0.69}_{-0.47}$ \\
080507 & L & ILT J153442.14+562609.0 & --- & $11.75\pm3.66$ & $2.13^{+1.48}_{-1.26}$ & $5.5^{+12.2}_{-3.3}$ & --- & --- & $1.05^{+1.15}_{-0.67}$ \\
140808A & L & ILT J144453.38+491305.9 & $3.290^{7}$ & $13.87\pm3.78$ & $2.14^{+0.08}_{-0.81}$ & $6.5^{+6.8}_{-1.9}$ & $105.6\pm28.8$ & $16.27^{+0.64}_{-6.15}$ & $1.11^{+0.89}_{-0.60}$ \\
191101A & L & ILT J164720.21+434437.0 & $0.681^{8}$ & $13.13\pm1.81$ & $0.46^{+1.07}_{-0.45}$ & $28.5^{+1470}_{-21.1}$ & $92.7\pm12.8$ & $3.22^{+7.54}_{-3.18}$ & $1.29^{+0.62}_{-0.45}$ \\
080916B & L & ILT J105437.13+690416.8 & --- & $24.27\pm6.64$ & $7.15^{+5.48}_{-4.09}$ & $3.4^{+6.7}_{-2.0}$ & --- & --- & $2.08^{+3.42}_{-1.26}$ \\
\hline 
050509B & S & ILT J123613.00+285902.9 & $0.225^{\rm 9}$ & $10.14\pm5.62$ & $2.04^{+0.35}_{-0.33}$ & $5.0^{+4.2}_{-3.1}$ & $36.3\pm20.1$ & $7.32^{+1.26}_{-1.17}$ & $0.11^{+0.15}_{-0.10}$ \\
050509B & S & ILT J123612.26+285929.2 & $0.225^{\rm 9}$ & $32.64\pm5.48$ & $2.99^{+0.14}_{-0.14}$ & $10.9^{+2.5}_{-2.2}$ & $116.9\pm19.6$ & $10.71^{+0.50}_{-0.49}$ & $0.69^{+0.30}_{-0.25}$ \\
    \hline
\end{tabular}
\begin{tablenotes}
\item \textit{Redshift sources:} $^1$\citet{Giarratana22}; $^2$\citet{Berger06}; $^3$SDSS \texttt{Photoz} table \citep{Beck16}; $^4$Galaxy clusters optical catalogue from SDSS DR6 \citep{Szabo11}; $^5$\citet{Jakobsson07}; $^6$\citet{Prochaska06}; $^7$\citet{Gorosabel14}; $^8$SDSS Quasar Catalogue \citep{Lyke20}; $^9$\citet{Prochaska05}.
\end{tablenotes}
\label{tab:crossmatch}
\end{threeparttable}
\end{table*}

\section{Host properties}
\label{sec:host_properties}

In this section, we examine the behaviour of our putative GRB hosts to assist with classification and examine how their properties compare to galaxy samples.

\subsection{Counterparts in other radio surveys}

To expand the frequency space for our putative hosts, we also performed crossmatches to other radio surveys. This allows us to better examine the spectral behaviour of the sources, in particular, allowing possible identification and classification of any AGN activity.

We crossmatched our putative hosts to the Westerbork Northern Sky Survey \citep[WENSS,][]{Rengelink97}, Faint Images of the Radio Sky at Twenty-centimeters \citep[FIRST,][]{Becker95,White97,Helfand15}, the NRAO Very Large Array Sky Survey \citep[NVSS,][]{Condon98}, Giant Metrewave Radio Telescope (GMRT) 150 MHz Survey \citep{Intema17}  and the Very Large Array Sky Survey \citep[VLASS,][]{Gordon21}. This extended our frequency regime to between 144 MHz and 3 GHz. We performed this crossmatch using VizieR\footnote{\url{https://doi.org/10.26093/cds/vizier}}, selecting the nearest source in each table within 20" to maximise the number of plausible matches. However, only five of our LoTSS sources had counterparts across these catalogues, given in Table \ref{tab:counterparts}.

In addition to the results in the survey catalogues, the host of GRB 200716C was also examined at radio wavelengths by \citet{Giarratana22}, who identified flux densities in the archival data of these surveys. We include their results when calculating the properties of this galaxy.

\begin{table*}
\centering
\caption{The candidate counterparts to our putative hosts in other radio surveys. The name prefixes are given in the brackets. Sources are ordered by \pchance~and GRB class.}
\begin{tabular}{ccccccc}
    \hline
    GRB & LoTSS match & WENSS & FIRST & NVSS & GMRT 150 MHz & VLASS \\
    & & (WENSS) & (FIRST) & (NVSS) & (TGSSADR) & (VLASS1QLCIR) \\
    \hline
201229A & ILT J140245.38+481150.7 & --- & J140245.4+481150 & --- & --- & --- \\
190211A & ILT J130635.93+415811.3 & B1304.3+4214 & J130635.9+415811 & J130635+415812 & J130635.9+415812 & J130635.94+415811.3 \\
220412A & ILT J081553.01+302035.9 & --- & J081552.9+302036 & --- & --- & --- \\
191101A & ILT J164720.21+434437.0 & --- & --- & J164721+434440 & --- & --- \\
    \hline
050509B & ILT J123612.26+285929.2 & --- & --- & J123612+285930 & --- & --- \\
\hline
\end{tabular}
\label{tab:counterparts}
\end{table*}

\subsection{Radio Spectral behaviour}

\begin{table*}
\centering
\caption{The spectral indices and \textit{WISE} colours and classifications of our putative hosts ordered by GRB class and \pchance. We use the criteria of \citet{Mingo16} and \citet{Mingo19} to determine classification. *s indicate sources where the colours are not fully constrained or that have redshifts greater than one and therefore the classifications are not fully constrained; and $\dagger$s indicate sources suggested to be AGN in other studies.}
\renewcommand{\arraystretch}{1.4}
\begin{tabular}{ccccccc}
    \hline
    GRB & LoTSS match & \alphalotss & \alphawide & $W1-W2$ & $W2 - W3$ & Implied \textit{WISE} class \\
    \hline
200716C & ILT J130402.61+293839.3 & $0.89 \pm 0.25$ & $-0.65 \pm 0.07$ & $0.13 \pm 0.11$ & $2.93 \pm 0.46$ & Star-forming \\
201229A & ILT J140245.38+481150.7 & $0.64 \pm 2.73$ & $0.43 \pm 0.07$  & $0.82 \pm 0.30$ & $<3.65$ & ULIRG/AGN*$\dagger$ \\
190211A & ILT J130635.93+415811.3 & $0.27 \pm 0.18$ & $-0.56 \pm 0.03$ & $0.53 \pm 0.33$ & $<4.38$ & ULIRG/AGN* \\
060123 & ILT J115847.46+453050.9 & --- & --- & $0.04 \pm 0.20$ & $4.23 \pm 0.35$ & ULIRG \\
091130B & ILT J133235.11+340524.3 & $2.35 \pm 1.59$ & --- & $0.34 \pm 0.08$ & $3.44 \pm 0.19$ & ULIRG* \\
110903A & ILT J130815.16+585857.2 & --- & --- & $0.31 \pm 0.14$ & $3.92 \pm 0.26$ & ULIRG \\
081025 & ILT J162131.13+602837.1 A & $0.14 \pm 1.24$ & --- & $-0.29 \pm 0.46$ & $<4.12$ & ULIRG* \\
150213B & ILT J165348.36+341126.2 & --- & --- & $0.10 \pm 0.35$ & $<4.15$ & ULIRG* \\
220412A & ILT J081553.01+302035.9 & --- & $0.02 \pm 0.01$ & $0.14 \pm 0.13$ & $<2.69$ & Star-forming* \\
081025 & ILT J162131.13+602837.1 B & $0.10 \pm 1.08$ & --- & $0.17 \pm 0.06$ & $<1.03$ & Elliptical \\
071020 & ILT J075839.09+325133.9 & --- & --- & $0.32 \pm 0.12$ & $3.51 \pm 0.31$ & ULIRG* \\
110521A & ILT J080031.58+454950.2 & --- & --- & $0.62 \pm 0.33$ & $<3.74$ & ULIRG/AGN* \\
060206 & ILT J133144.19+350305.6 & --- & --- & $0.28 \pm 0.11$ & $2.66 \pm 0.54$ & Star-forming \\
080507 & ILT J153442.14+562609.0 & --- & --- & $0.57 \pm 0.14$ & $<3.13$ & AGN* \\
140808A & ILT J144453.38+491305.9 & --- & --- & $1.05 \pm 0.28$ & $<4.04$ & ULIRG/AGN* \\
191101A & ILT J164720.21+434437.0 & --- & $0.58 \pm 0.31$ & $0.80 \pm 0.13$ & $2.75 \pm 0.12$ & AGN$\dagger$ \\
080916B & ILT J105437.13+690416.8 & --- & --- & $0.42 \pm 0.09$ & $4.13 \pm 0.14$ & ULIRG* \\
\hline
050509B & ILT J123613.00+285902.9 & $-0.21 \pm 0.24$ & --- & $0.27 \pm 0.06$ & $<1.29$ & Elliptical \\
050509B & ILT J123612.26+285929.2 & $0.27 \pm 0.41$ & $-0.58 \pm 0.18$ & $<0.91$ & $<3.87$ & ULIRG/AGN* \\
\hline
\end{tabular}
\label{tab:indices}
\end{table*}

Following \citet{Shimwell22}, the LoTSS band can be divided into three 16 MHz wide bands centred at 128, 144 and 160 MHz. We downloaded the primary beam corrected images for each of these bands and extracted the flux densities for our putative hosts using the \textsc{PyBDSF}\footnote{\url{https://www.astron.nl/citt/pybdsf/index.html}} package, setting both the \texttt{thresh\_pix} and \texttt{thresh\_isl} parameters to 3 (i.e. a detection significance of  3-$\sigma$). The flux densities were then fitted with a power law model, $F_{\nu} = k \nu^{\alpha}$, to derive spectral indices, given in Table \ref{tab:indices} as \alphalotss. The varying significance of each source across the LoTSS bandwidth means that the source was not necessarily detected in each band and therefore we could only derive spectral indices for seven sources. We attempted to increase this number by lowering the significance thresholds but found that the resultant spectral indices were unphysically extreme, likely indicative of noise being extracted rather than real sources. In general, we found our values for \alphalotss~to be poorly constrained, likely due to large errors in the flux density measurements and the relatively narrow frequency range. We note, however, that this is consistent with the findings of \citet{Shimwell22}.

For the sources with counterparts at GHz frequencies, we refitted the power law model but included both the flux densities given in those catalogues and the flux density across the full LoTSS band centred at 144 MHz. The resulting spectral indices are given in Table \ref{tab:indices} as \alphawide~and our sources' fitted SEDs are shown in Figure \ref{fig:seds}.

For most sources with both \alphalotss~and \alphawide, we found that both values for $\alpha$ are reasonably consistent. However, in the cases of ILT J123612.26+285929.2 and ILT J130402.61+293839.3, the \alphalotss~and \alphawide~indicate opposite behaviour for the spectrum. In this case, as with other sources with both \alphalotss~and \alphawide, we are inclined to trust \alphawide~more due to \alphalotss~being poorly constrained, as mentioned above. It is also plausible that the true SED has a more complex structure than the single power law assumed and, for instance, \alphalotss~and \alphawide~capture separate parts of a broken power law. Measurements with the LOFAR low-band antennas (LBA) may be useful in further constraining any spectral turnover.

\subsection{IR behaviour}
\label{sec:IR}

\begin{figure}
    \centering
    \includegraphics[width=\columnwidth]{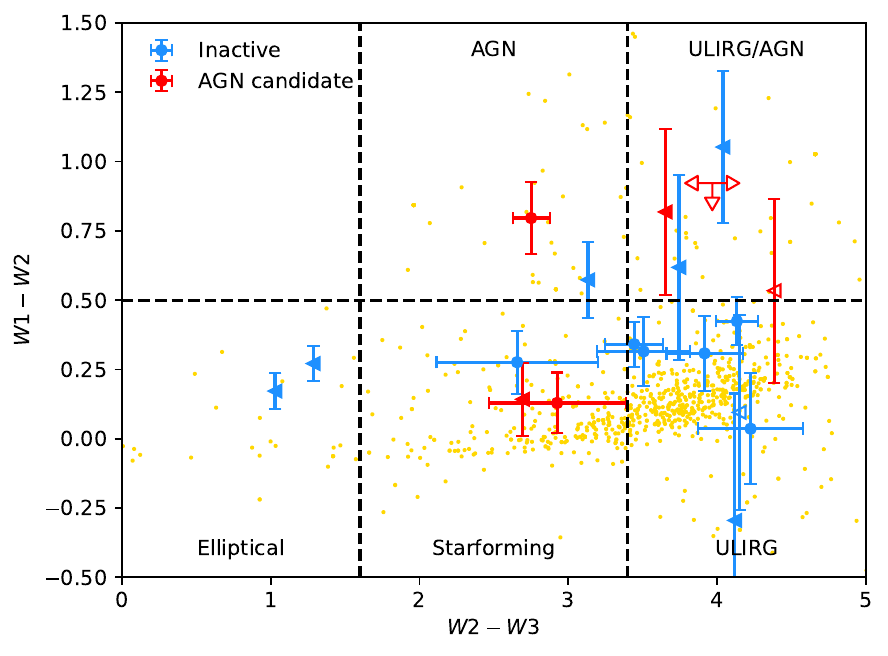}
    \caption{The $W1-W2$ and $W2-W3$ colours of the closest AllWISE sources to our putative host sample. Our CCSNe comparison sample is plotted in gold. Red points indicate putative GRB hosts with AGN indicators either from literature or from their spectral indices while blue points indicate sources with no other traces of AGN activity. A $\blacktriangleleft$ indicates an upper limit in $W2-W3$ and the triple arrow indicates $W1-W2$ is an upper limit and $W2-W3$ is unconstrained. Open points indicate sources with a separation >5" between the AllWISE and the LoTSS source.}
    \label{fig:wise_colours}
\end{figure}

We also examined the infrared colours of the galaxies in the \textit{Wide-field Infrared Survey Explorer} (\textit{WISE}) bands. These colours are widely used to assist with classifying galaxies and help evaluate any AGN activity. We matched our sources to the closest objects in the AllWISE catalogue \citep{allwise} finding that all such matches had a \pchance<1\%. We used the criteria of \citet{Mingo16} and \citet{Mingo19}, to broadly classify galaxies into AGN, luminous IR galaxies (LIRGs), ultra-luminous IR galaxies (ULIRGs), star-forming and elliptical classes. We tabulate the $W1-W2$ and $W2-W3$ colours in Table \ref{tab:indices} and plot them in Figure \ref{fig:wise_colours}.

We also derived \textit{WISE} colours for the CCSNe sample, also shown in Figure \ref{fig:wise_colours}. We found that all types of CCSNe follow similar behaviour with no significant differences between them. This behaviour is also compatible with the distribution exhibited by our GRB host candidates although the sources matched to GRBs are more likely to have colours consistent with AGN activity.  We further examine such activity in the following section.

According to our criteria, both the GRB and CCNSe samples are dominated by ULIRGs. While it might be expected that a greater proportion of each sample would be star-forming galaxies, LIRGs and ULIRGs are common radio-selected hosts for these transients \citep[e.g.][]{Perley15}. The nature of these galaxies enhances star formation, with the star formation efficiency increased by a factor of 2 -- 3 compared to the general galaxy population. During an extreme starburst, this enhancement can reach up to an order of magnitude \citep{GraciaCarpio08,GarciaBurillo12}. Other radio studies of GRB hosts have found them to be consistent with LIRGs or ULIRGs and identified high SFRs of order 50 -- 200 \sfrunits~\citep{Perley15}. These galaxies were also found to be significantly lower luminosity relative to their SFRs than field galaxies, a result consistent with other studies \citep[e.g.][]{Levesque14,Perley16,Lyman17}. This is compatible with the suggestion that there is a metallicity bias or cutoff in long GRB hosts. These high SFRs and low metallicities are ideal for creating an environment rich with collapsars, the progenitors of both long GRBs and CCSNe. There could also be systematic reasons why ULIRGs are so dominant. It is likely that our criteria over simplify the complex behaviour of the galaxies and therefore neglect the significant overlap between different classifications. In addition, as discussed in Section \ref{sec:selection}, there are significant selection effects inherent to a radio search. This biases towards selecting a different population of galaxies than optical surveys for instance, and such a sample would be expected to include a greater proportion of ULIRGs than the overall population of galaxies. 

\subsection{Star formation}

\begin{table*}
\centering
\caption{The masses and SFRs of our putative hosts. The masses are taken from SDSS. The SFRs are calculated using the prescriptions of \citeauthor{Gurkan18} ($\psi_{\rm G}$, Equation \ref{eq:SFR_Gurkan}), \citeauthor{Smith21} ($\psi_{\rm S}$, Equation \ref{eq:SFR_Smith}) and \citeauthor{Bell03} ($\psi_{\rm B}$, Equation \ref{eq:SFR_Bell}). 
}
\renewcommand{\arraystretch}{1.4}
\begin{threeparttable}
\begin{tabular}{ccccccc}
    \hline
    GRB & LoTSS match & Redshift & $\log(M_{\rm gal} / M_{\odot})$ & $\psi_{\rm G}$ & $\psi_{\rm S}$ & $\psi_{\rm B}$ \\
    & & & & (\sfrunits) & (\sfrunits) & (\sfrunits)\\
    \hline
200716C & ILT J130402.61+293839.3 &  0.341 & --- & $111.1^{+11.9}_{-10.7}$ & --- & $167\pm21.2$ / $338.3\pm84.6$${\dagger}$ \\
060123 & ILT J115847.46+453050.9 &  0.562 & --- & $67.0^{+23.4}_{-20.9}$ & --- & --- \\
091130B & ILT J133235.11+340524.3 &  0.282 & --- & $26.0^{+3.8}_{-3.5}$ & --- & --- \\
110903A & ILT J130815.16+585857.2 &  0.311 & --- & $11.4^{+3.1}_{-2.9}$ & --- & --- \\
150213B & ILT J165348.36+341126.2 &  0.704 & --- & $96.9^{+32.6}_{-29.0}$ & --- & --- \\
220412A & ILT J081553.01+302035.9 &  0.402 & --- & $43.1^{+12.0}_{-10.8}$ & --- & $379.8\pm44.8$ \\
071020 & ILT J075839.09+325133.9 &  2.146 & $<9.43$ & $1082.7^{+432.2}_{-369.6}$ & $>3999.6$ & --- \\
060206 & ILT J133144.19+350305.6 &  4.050 & --- & $6276.5^{+2233.7}_{-1862.7}$ & --- & --- \\
140808A & ILT J144453.38+491305.9 &  3.290 & --- & $5610.7^{+1323.5}_{-1115.9}$ & --- & --- \\
191101A & ILT J164720.21+434437.0 &  0.681 & --- & $112.0^{+26.7}_{-23.6}$ & --- & $4597.7\pm560.7$ \\
    \hline
050509B & ILT J123613.00+285902.9 &  0.225 & $ 11.08^{+0.03}_{-0.03}$ & $79.9^{+9.5}_{-8.5}$ & $48.7^{+16.3}_{-10.0}$ & --- \\
050509B & ILT J123612.26+285929.2 &  0.225 & --- & $143.2^{+14.5}_{-13.0}$ & --- & $316.4\pm40.6$ \\
\hline
\end{tabular}
\begin{tablenotes}
\item ${\dagger}$\citet{Giarratana22} find significantly different flux densities in FIRST and NVSS. Here, the first value uses the FIRST flux density and the second uses the NVSS flux density. 
\end{tablenotes}
\end{threeparttable}
\label{tab:sfrs}
\end{table*}

The star formation history of a galaxy has a significant effect on the rate of GRBs that take place within it. In particular, long GRB progenitors are massive and short-lived stars so the hosts of such GRBs are generally found to have significantly higher SFRs than the majority of the field galaxy population. Here, we examine these properties of our putative hosts.

\subsubsection{Star-forming galaxies}

The spectral index of a galaxy can be indicative of whether it is star-forming and we therefore compare our measured spectral indices to those expected of star-forming galaxies. The radio SEDs of such objects are typically expected to be a superposition of two power laws, a thermal and a non-thermal component \citep[e.g.][]{Duric88}:
\begin{equation}
    S_{\rm tot}(\nu) = S_{\rm th}(\nu_0)\left(\frac{\nu}{\nu_0}\right)^{-0.1} + S_{\rm nth}(\nu_0)\left(\frac{\nu}{\nu_0}\right)^{\alpha_{\rm nth}} 
\end{equation}
The thermal fraction is estimated to be $f_{\rm th}\sim10$\% \citep{Niklas97,Tabatabaei17} and hence our measured \alphalotss~and \alphawide~should be comparable or slightly smaller in absolute terms than $\alpha_{\rm nth}$ \citep{Tabatabaei17}.

The radio spectral index of star-forming galaxies is typically thought to be $\alpha_{\rm nth}\sim-1.0$ for the non-thermal index and $\sim-0.8$ for the total index \citep{Tabatabaei17}. However, this simple picture of superimposed power laws does not appear consistent with observations at lower frequencies. For instance, the sample of galaxies examined by \citet{Klein18} displayed a broken or exponentially declining power law spectrum, with the break or decline occurring between 1--12 GHz, as $\alpha_{\rm nth}$ varies with frequency. The resultant power law at lower frequencies is significantly shallower than at higher frequencies, with a spectral index of $\sim-0.6$. This behaviour has also been identified in studies that reach even lower to MHz frequencies \citep{Yun02,CalistroRivera17,An21}. 

From the spectral indices in Table \ref{tab:indices}, the most likely candidates for star-forming galaxies are ILT J130402.61+293839.3, ILT J130635.93+415811.3 and ILT J123612.26+285929.2. From their \textit{WISE} colours, only ILT J130402.61+293839.3 was consistent with being star-forming while the other two sources had been categorised as ULIRG/AGN, although high star formation is still likely to be present as discussed above. There were two other sources classified as star-forming by their \textit{WISE} colours, ILT J081553.01+302035.9 and ILT J133144.19+350305.6. Unfortunately, these sources' spectral indices are either unavailable or poorly constrained due to a low number of data points. It is therefore difficult to determine if the radio evidence independently points towards them being star-forming.

\subsubsection{Star formation rates with LoTSS flux densities}

To derive SFRs for our matched sources with redshifts, we returned to Equations \ref{eq:SFR_Gurkan} and \ref{eq:SFR_Smith}. As noted above, Equation \ref{eq:SFR_Smith} requires the mass of the host to be known which is only the case for ILT J123613.00+285902.9 \citep{Savaglio09}, and an upper limit for ILT J075839.09+325133.9 \citep{Perley16}. We were therefore able to derive 12 SFRs using Equation \ref{eq:SFR_Gurkan} and two using Equation \ref{eq:SFR_Smith}. Where we have both estimates, we find they are generally comparable. The remaining discrepancies may due to the sensitivity of Equation \ref{eq:SFR_Smith} to the galaxy mass and therefore the accuracy of its measurement.

We found that three of our matched sources (ILT J075839.09+325133.9, ILT J133144.19+350305.6 and ILT J144453.38+491305.9), in addition to a significant proportion of the general population, exhibit apparent extreme star formation rates of several thousand \sfrunits~or greater. There are examples of such high values in previous work, such as the estimate of SFR $>$ 1000 \sfrunits~for the high redshift GRB 090404 \citep{Perley17}. There is also a possibility of afterglow contamination, as suggested for the case of GRB 100814A \citep{Stanway14}, but it is unlikely that these sources are significantly affected. Such high values could indicate incorrectly assigned redshifts rather than the hosts, or significant contributions from other emission sources. Recent work using LoTSS to examine the cosmic star formation history showed that significant scatter in the relationship between $L_{150}$ and SFR is likely the result of AGN \citep{Cochrane23}. It is therefore probable that these apparently extreme SFRs are actually AGN rather than star formation emission.

To provide a comparison, we also derived SFRs for the CCSNe sample and the \textsc{InactiveField} sample using Equation \ref{eq:SFR_Gurkan}. To account for nearby resolved CCSNe host galaxies, we summed the SFRs of all LoTSS counterparts for each supernova. To compare the GRB hosts to our CCSNe and general sample, we set an upper limit on the SFR of 200 \sfrunits. This limits our sample to the SFRs most likely to be real for both for the GRB hosts and the comparison samples. Excluding the possible hosts of the short GRB 050509B, we found that the long GRB hosts have a mean SFR of $66.8 \pm 38.2$ \sfrunits, while the general inactive galaxy population has a mean SFR of $18.2 \pm 33.1$ \sfrunits. It is clear that our putative long GRB hosts have significantly greater apparent SFRs than most galaxies in the field, although AGN contamination could still contribute to apparently enhanced radio flux, a factor we address in Section \ref{sec:agn}. We note also that the \textsc{InactiveField}'s inferred SFR distribution is somewhat higher than might be expected for a general galaxy population, which may be indicative of incorrect crossmatching, redshift assignments or lingering AGN contamination. The CCSNe hosts have even lower SFRs than either of the other two samples at $8.5 \pm 23.3$ \sfrunits, possibly due to their small redshifts, and therefore their low luminosities and being resolvable into multiple components. This low SFR behaviour was again common to all types of CCSNe.

The results for the matches to the short GRB 050509B are more surprising, specifically ILT J123613.00+285902.9. For this source, both our methods indicate SFRs of tens of \sfrunits, much greater than the UV measured $<0.2$ \sfrunits~of \citet{Gehrels05}. While dust obscuration is likely to cause some discrepancy between measurements in these two regimes, it is implausible that two orders of magnitude difference is possible. If the radio is, indeed, dominated by star formation, it is therefore likely that ILT J123613.00+285902.9 is not the radio counterpart to the optically-assigned large elliptical galaxy 2MASX J12361286+285858026 despite their spatial consistency. Alternatively, it is possible that this source is dominated by AGN emission and we discuss this further below.

\subsubsection{Star formation rates with flux densities from other radio surveys}

For those sources with higher frequency counterparts, we can also use the prescription of \citet{Bell03} to derive SFRs, similarly to \citet{Michalowski09}. Using the luminosity of the source at 1.4 GHz, $L_{1.4}$, in W Hz$^{-1}$, the SFR is found to be 
\begin{equation}
\label{eq:SFR_Bell}
    \psi_{\rm B} = \begin{cases} 5.52 \times 10^{-22} L_{1.4}, & L_{1.4} > L_c \\
    \frac{5.52 \times 10^{-22}}{(0.1 + 0.9 (L_{1.4} / L_c)^{0.3}}L_{1.4}, & L_{1.4} \leq L_c
    \end{cases}
\end{equation}
where $L_c = 6.4\times 10^{21}$ W Hz$^{-1}$ is some critical 1.4 GHz luminosity.

Four sources have both higher frequency counterparts and associated redshifts allowing us to evaluate their SFRs with Equation \ref{eq:SFR_Bell}. We found that there was reasonable agreement (a factor of a few) for two of these sources, ILT J130402.61+293839.3 and ILT J123612.26+285929.2, but there were much more significant discrepancies for ILT J081553.01+302035.9 (a factor of $\sim9$) and ILT J164720.21+434437.0 (a factor of $\sim40$). We note that the spectrum of J081553.01+302035.9 is significantly flatter than a typical star-forming galaxy, but other evidence, such as the \textit{WISE} colours do point towards this source as being star-forming. It is therefore unclear why such a large discrepancy is present. ILT J164720.21+434437.0, on the other hand, is an AGN candidate and it is likely that the higher frequency flux density and therefore this SFR measurement is actually dominated by AGN activity rather than star formation.

\subsection{AGN contamination}
\label{sec:agn}

A large proportion of the radio source population is made up of active galaxies and as such they represent a significant possible contaminant in our sample of putative hosts. We therefore investigated our sample to determine whether any sources were likely to actually be AGN and therefore most probably unrelated to the GRBs.

\subsubsection{Literature candidates}

LoTSS has previously been extensively investigated for AGN and we initially compared our source list to the AGN catalogues for HETDEX Spring Field of LoTSS DR1 \citep{Williams19,Hardcastle19,Mingo19} and LoTSS Deep Fields Data Release 1 \citep{Best23}. No crossmatches were identified and we therefore also performed a wider literature search. This identified ILT J164720.21+434437.0 \citep{Lyke20} and ILT J140245.38+481150.7 \citep{Truebenbach17}, the putative hosts of GRB 191101A and GRB 201229A respectively, as having possible AGN counterparts. While the separation of GRB 191101A indicates that this is unlikely to be an accurate match, the crossmatch between GRB 201229A and ILT J140245.38+481150.7 inspires much more confidence. \citet{Truebenbach17} suggest that ILT J140245.38+481150.7 is an AGN due to a combination of radio/IR emission but a lack of optical detections and find such methods to be consistent with other AGN selection criteria. However, when we perform similar analysis below, we find that it differs in its behaviour from the vast majority of AGN while its \textit{WISE} colours are consistent with both AGN and ULIRG classifications. If this were an AGN, the small separation could mean that this is the first long GRB to be found to be associated with an active galaxy.

\subsubsection{Radio behaviour}

The morphology of a radio source can indicate the presence of an AGN. In particular, irregular morphologies or the physical size of a source can be the result of activity. This extends to the LoTSS frequencies, as shown for Fanaroff-Riley class galaxies in the LoTSS-Deep field \citep{Mingo22}. None of our sources have physical extents comparable to \citeauthor{Mingo22}'s sample, however, and in general our sources' morphologies are significantly more regular. There a few exceptions, namely ILT J162131.13+602837.1, ILT J10543713+690416.8 and ILT J123613.00+285902.9. We have already determined that ILT J162131.13+602837.1's morphology is the result of two galaxies being close enough to be unresolved in LoTSS DR2, however, the other sources are both plausibly AGN dominated. In particular, the irregular and elongated morphology of ILT J123613.00+285902.9 is consistent with the presence of a radio jet. Elliptical galaxies are not uncommon hosts for low power radio galaxies and it is possible that 2MASX J12361286+285858026 is the host of ILT J123613.00+285902.9. However, there are other optical sources spatially consistent with ILT J123613.00+285902.9 as shown by \citet{Hjorth05}. The middle panel of their Figure 1 shows the field with 2MASX J12361286+285858026 subtracted out and a new source identified to its North. While \citeauthor{Hjorth05} do not examine this source in greater detail and it is unclear whether it is foreground or background, we encourage further investigations to determine whether it is the true counterpart to ILT J123613.00+285902.9.

The radio spectrum of a galactic source can also be a clue as to its activity. The canonical radio spectral index for AGN is typically taken to be $\sim -0.7$ \citep{Condon02} and the low-frequency samples examined by \citet{CalistroRivera17} and \citet{Gurkan18} in addition to the AGN sample in LoTSS DR1 appeared consistent with this value \citep[Figure 3 of][]{Sabater19}. However, as the flux density of an AGN at 1.4 GHz decreased, its spectral index became significantly shallower. Four of our sources with 1.4 GHz detections have indices consistent with the canonical value (ILT J130402.61+293839.3, ILT J130635.93+415811.3, ILT J081553.01+302035.9 and ILT J123612.26+285929.2). As previously mentioned, ILT J164720.21+434437.0's spectral index is sensitive to the fit and it is plausible that this is also consistent. The poorly constrained nature of \alphalotss~means we cannot firmly determine whether any of the remaining  sources are also consistent.

However, a great deal of diversity around the canonical spectral index has been observed across the AGN population. For instance, peaked spectrum (PS) sources have spectra with distinct peaks and steep drops around them \citep[e.g.][]{ODea91,ODea21}, possibly as the result of synchrotron self-absorption or free-free absorption \citep{Bicknell97}. Such sources include GHz peaking sources (GPS) and compact steep spectrum (CSS) sources which peak at lower frequencies of a few hundred MHz. This means that in the regime probed by LOFAR and the catalogued data, they can actually display positive (by our convention) or flat spectral indices \citep[e.g.][]{ODea98,Sadler16}. The turnover suggested in the SED of ILT J130635.93+415811.3 could be due to it being such a CSS source. There are other classes of AGN that could display diverging spectral behaviour, such as AGN with emission primarily arising from advection dominated accretion flows \citep[ADAFs, e.g.][]{Narayan94,Narayan98,Mahadevan98}. Finally, low luminosity jets can also result in flat spectra \citep[e.g.][]{Falcke99}. We also note that the canonical spectral index for AGN is very similar to that of starforming galaxies \citep[e.g.][]{CalistroRivera17}, further diluting spectral behaviour as an indicator of AGN behaviour.

It is therefore difficult to determine which sources could be AGN solely from their spectral indices and the majority of our host candidates display behaviour which could be consistent with an AGN origin. We therefore return to the infrared behaviour of the sources in the following section.

\subsubsection{\textit{WISE} colours and \qir}

The \textit{WISE} colours detailed in Section \ref{sec:IR} can be an indicator of AGN and seven of the sources in our sample were found to have colours consistent with AGN classification. However, only one source (ILT J164720.21+434437.0) is sufficiently constrained to ensure this classification. We note also that the $W1 - W2$ criterion does not necessarily confirm a lack of AGN behaviour, as some classes of AGN such as Seyferts may lie below it. There are also significant evolutionary effects with redshift that can greatly change a source's position on this diagram \citep{Assef13}. It is therefore possible that sources with redshifts of $z>1$ are misclassified. For our sample, this includes ILT J075839.09+325133.9, ILT J133144.19+350305.6 and ILT J144453.38+491305.9. We therefore investigate further criteria to more fully examine possible AGN contamination.

The ratio of IR to radio emission can also indicate the presence of AGN behaviour \citep[e.g.][]{Helou85,Condon91},
\begin{equation}
    \label{eq:qir}
    q_{\rm IR} = \log\frac{S_{\rm IR}}{S_{\rm radio}}
\end{equation}
where $S$ is the flux density in a given filter or at a given frequency. Radio AGN generally have significantly lower values of \qir~than star forming galaxies which also display an evolution with redshift \citep[e.g.][]{CalistroRivera17}. While typically longer wavelengths and higher frequencies are employed for $S_{\rm IR}$ and $S_{\rm radio}$ respectively, a significant AGN correlation could still be expected for the \textit{WISE} bands and 144 MHz LoTSS frequency we have available. We calculated \qir~for each source for each available \textit{WISE} band counterpart and show these in Table \ref{tab:qir}.

\begin{table*}
\centering
\caption{The log ratio of IR to radio emission, \qir, for each \textit{WISE} band of our putative hosts ordered by GRB class and \pchance.}
\renewcommand{\arraystretch}{1.4}
\begin{tabular}{cccccc}
    \hline
    GRB & LoTSS match & $q_{W1}$ & $q_{W2}$ & $q_{W3}$ & $q_{W4}$ \\
    \hline
200716C & ILT J130402.61+293839.3 & $-0.19^{+0.03}_{+0.03}$ & $-0.14^{+0.05}_{+0.05}$ & $1.03^{+0.17}_{+0.17}$ & --- \\
201229A & ILT J140245.38+481150.7 & $-0.36^{+0.10}_{+0.09}$ & $-0.04^{+0.13}_{+0.12}$ & --- & --- \\
190211A & ILT J130635.93+415811.3 & $-2.29^{+0.04}_{+0.04}$ & $-2.08^{+0.09}_{+0.09}$ & --- & --- \\
060123 & ILT J115847.46+453050.9 & $0.23^{+0.17}_{+0.13}$ & $0.24^{+0.21}_{+0.17}$ & $1.93^{+0.23}_{+0.19}$ & --- \\
091130B & ILT J133235.11+340524.3 & $0.38^{+0.06}_{+0.05}$ & $0.51^{+0.06}_{+0.06}$ & $1.89^{+0.10}_{+0.10}$ & $2.88^{+0.19}_{+0.18}$ \\
110903A & ILT J130815.16+585857.2 & $0.46^{+0.13}_{+0.11}$ & $0.58^{+0.15}_{+0.13}$ & $2.15^{+0.18}_{+0.16}$ & --- \\
081025 & ILT J162131.13+602837.1 A & $-1.09^{+0.06}_{+0.06}$ & $-1.21^{+0.18}_{+0.17}$ & --- & --- \\
150213B & ILT J165348.36+341126.2 & $-0.06^{+0.17}_{+0.14}$ & $-0.02^{+0.24}_{+0.21}$ & --- & --- \\
220412A & ILT J081553.01+302035.9 & $0.37^{+0.12}_{+0.10}$ & $0.43^{+0.15}_{+0.12}$ & --- & --- \\
081025 & ILT J162131.13+602837.1 B & $-0.36^{+0.03}_{+0.03}$ & $-0.29^{+0.03}_{+0.03}$ & --- & --- \\
071020 & ILT J075839.09+325133.9 & $0.66^{+0.17}_{+0.13}$ & $0.78^{+0.19}_{+0.15}$ & $2.18^{+0.25}_{+0.20}$ & --- \\
110521A & ILT J080031.58+454950.2 & $-0.43^{+0.17}_{+0.14}$ & $-0.18^{+0.21}_{+0.18}$ & --- & --- \\
060206 & ILT J133144.19+350305.6 & $0.48^{+0.13}_{+0.11}$ & $0.59^{+0.15}_{+0.12}$ & $1.65^{+0.30}_{+0.28}$ & --- \\
080507 & ILT J153442.14+562609.0 & $0.04^{+0.13}_{+0.11}$ & $0.27^{+0.15}_{+0.12}$ & --- & --- \\
140808A & ILT J144453.38+491305.9 & $-0.45^{+0.10}_{+0.10}$ & $-0.03^{+0.12}_{+0.11}$ & --- & $2.76^{+0.26}_{+0.25}$ \\
191101A & ILT J164720.21+434437.0 & $0.81^{+0.12}_{+0.11}$ & $1.13^{+0.09}_{+0.08}$ & $2.23^{+0.12}_{+0.11}$ & $3.31^{+0.18}_{+0.17}$ \\
080916B & ILT J105437.13+690416.8 & $0.01^{+0.29}_{+0.18}$ & $0.18^{+0.30}_{+0.19}$ & $1.83^{+0.31}_{+0.20}$ & $3.00^{+0.34}_{+0.23}$ \\
\hline
050509B & ILT J123613.00+285902.9 & $0.02^{+0.03}_{+0.03}$ & $0.13^{+0.04}_{+0.04}$ & --- & --- \\
050509B & ILT J123612.26+285929.2 & $-1.69^{+0.09}_{+0.09}$ & --- & --- & --- \\
\hline
\end{tabular}
\label{tab:qir}
\end{table*}

\begin{figure}
    \centering
    \includegraphics[width=\columnwidth]{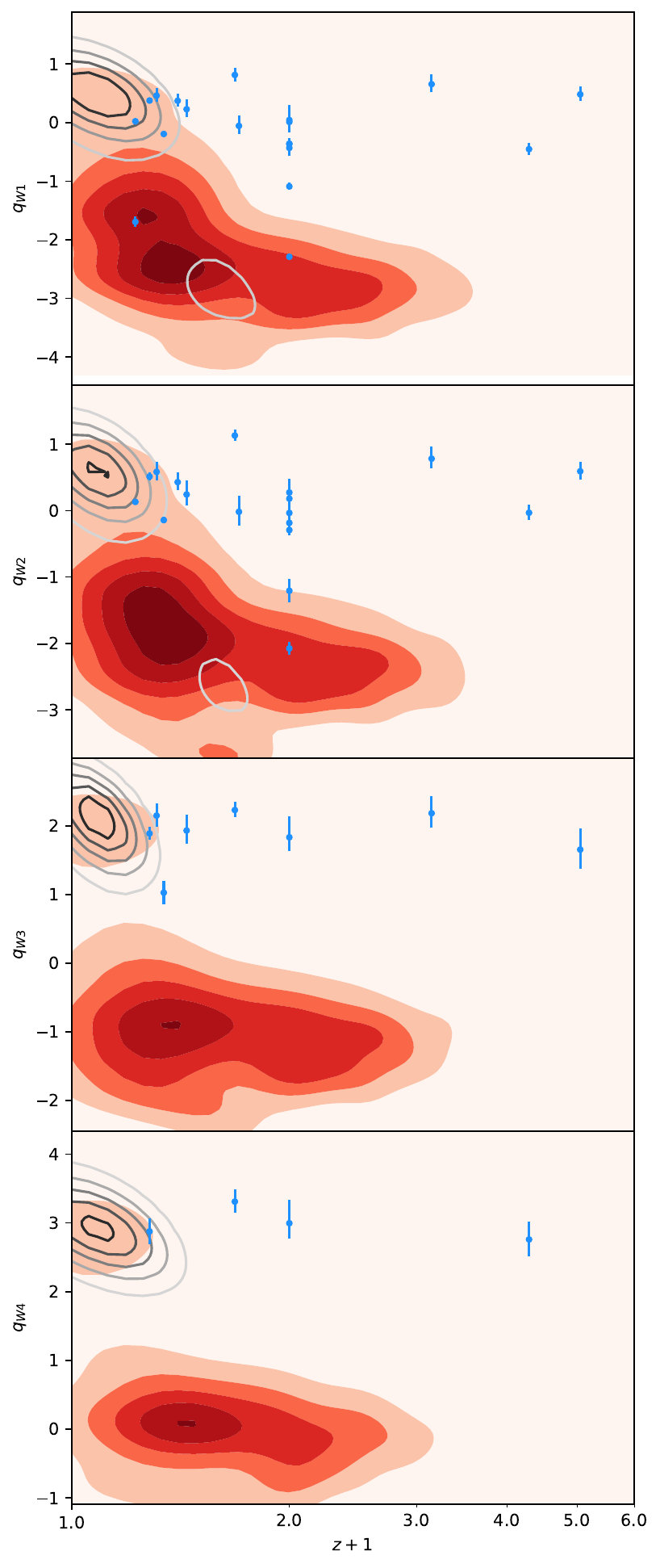}
    \caption{\qir~calculated using each \textit{WISE} band and the LoTSS 144 MHz flux densities. The red shaded regions indicate the distribution of, from top to bottom, 11102, 11003, 8104 and 5356 SIMBAD AGN, the greyscale contours indicate the distribution of apparently inactive field galaxies and the blue points indicate our putative GRB hosts. Note the red shading is log scale and the varying range of each panel's y axis.}
    \label{fig:q_all}
\end{figure}

We also calculated \qir~for the \textsc{ActiveField} and \textsc{InactiveField} samples and compare them to our matched sources in Figure \ref{fig:q_all} in the red shading and greyscale contours respectively. For our putative hosts without known redshifts, we assumed $z=1$. While the peak of the long GRB redshift distribution is $\sim2.2$ \citep{Evans09}, we choose a lower value to account for the inherent bias in a radio search towards lower redshift sources. We find that the majority of AGN do exhibit smaller values of \qir~than the field galaxies, although there is significant crossover and the redshift distribution of the field galaxies is much smaller than that of the AGN. There is also the possibility of incomplete or inaccurate classifications in SIMBAD. Nevertheless, it appears that most of our GRB host candidates exhibit different behaviour to the vast majority of the AGN sample. While this does not preclude any of them from being AGN, it does imply that they would be outliers to the main AGN population.

However, there are noticeable outliers to the other GRB hosts in the top two panels of Figure \ref{fig:q_all}. These are ILT J130635.93+415811.3, ILT J162131.13+602837.1 A and ILT J123612.26+285929.2 associated with GRBs 190211A, 081025 and 050509B respectively. There is significant evidence that both ILT J130635.93+415811.3 and ILT J123612.26+285929.2 are indeed AGN and their selection was due to the extreme flux densities induced by their active nature. However, it is more difficult to confirm if ILT J162131.13+602837.1 A is an AGN due to a lack of redshift constraint. However, from the measured fluence of GRB 081025, we expect $z\sim2.5$ for $E_{\gamma,{\rm iso}}\sim10^{53}$ erg, typical for a long GRB, or $z\sim0.4$ for $E_{\gamma,{\rm iso}}\sim10^{51}$ erg, if GRB 081025 is an extremely low luminosity GRB. While this lower redshift could place it towards the higher density AGN parameter space, the more probable higher redshift is inconsistent with the majority of AGN. We also note it is less of an outlier to our putative host distribution than the other two sources.

\section{Final candidate host sample}\label{sec:final_sample}

We now refine the initial list of associations given in Table \ref{tab:crossmatch} to define a final sample of GRBs and their hosts that are most likely accurate crossmatches.

From our original sample in Table \ref{tab:counterparts}, the large angular or physical separations between 10/19 possible associations (GRB 190211A/ILT J130635.93+415811.3, GRB 081025/ILT J162131.13+602837.1 B, GRB 071020/ILT J075839.09+325133.9, GRB 110521A/ILT J080031.58+454950.2, GRB 060206/ILT J133144.19+350305.6, GRB 140808A/ILT J144453.38+491305.9, GRB 191101A/ILT J164720.21+434437.0, GRB 080916B/ILT J105437.13+690416.8 and GRB 050509B/ILT J123612.26+285929.2) casts doubts on these being accurate counterparts. Two of these GRBs also have crossmatches at lower separations (GRB 081025/ILT J162131.13+602837.1 A and GRB 050509B/ILT J123613.00+285902.9) which are more likely to be the true counterpart.

In addition, a significant number of sources are likely to be AGN and therefore most likely unrelated to the GRBs. While we cannot necessarily ensure any of our crossmatches aren't subject to AGN contamination, 7/19 candidates (GRB 190211A/ILT J130635.93+415811.3, GRB 110521A/ILT J080031.58+454950.2, GRB 080507/ILT J153442.14+562609.0, GRB 140808A/ILT J144453.38+491305.9, GRB 191101A/ILT J164720.21+434437.0, GRB 050509B/ILT J123613.00+285902.9 and GRB 050509B/ILT J123612.26+285929.2) are most probably AGN related.

Eliminating these sources therefore leaves a final sample of eight matches, all for long GRBs, given in Table \ref{tab:final_sample}. We note that these criteria also eliminate the sources with extreme SFRs $\gtrsim200$ \sfrunits~implying that these were likely due to inaccurate crossmatching and therefore redshift assignments or that their radio flux is actually AGN dominated.

\begin{table*}
\centering
\caption{Our final sample of GRB and host crossmatch candidates ordered by \pchance. *s indicate sources where the \textit{WISE} colours and therefore classifications are not fully constrained.}
\renewcommand{\arraystretch}{1.4}
\begin{tabular}{cccccc}
    \hline
    GRB & LoTSS match & Redshift & Implied \textit{WISE} class & $\psi_{\rm G}$ & $\psi_{\rm B}$\\
    & & & & (\sfrunits) & (\sfrunits)\\
    \hline
200716C & ILT J130402.61+293839.3 & 0.341 & Starforming & $111.1^{+11.9}_{-10.7}$ & $167\pm21.2$ / $338.3\pm84.6$ \\
201229A & ILT J140245.38+481150.7 & --- & ULIRG/AGN* & --- & --- \\
060123 & ILT J115847.46+453050.9 & 0.442 & ULIRG & $39.0^{+13.3}_{-12.0}$ & --- \\
091130B & ILT J133235.11+340524.3 & 0.282 & ULIRG* & $26.0^{+3.8}_{-3.5}$ & --- \\
110903A & ILT J130815.16+585857.2 & 0.311 & ULIRG & $11.4^{+3.1}_{-2.9}$ & --- \\
081025 & ILT J162131.13+602837.1 A & --- & ULIRG* & --- & --- \\
150213B & ILT J165348.36+341126.2 & 0.704 & ULIRG* & $96.9^{+32.6}_{-29.0}$ & --- \\
220412A & ILT J081553.01+302035.9 & 0.402 & Starforming* & $43.1^{+12.0}_{-10.8}$ & $379.8\pm44.8$ \\
\hline
\end{tabular}
\label{tab:final_sample}
\end{table*}

The redshift distribution of this sample is low compared to the majority of the long GRB population. This is bias is consistent with that expected from selection effects, as discussed in Section \ref{sec:selection}. In terms of offsets, the angular offsets for the majority of this sample are less than 10", the only exception being between GRB 081025 and ILT J162131.13+602837.1 A at $\sim16$". These lead to normalised offsets of $\leq6.6$, consistent with the findings of \citet{Blanchard16}, while the mean physical offset is $31.3\pm15$ kpc. This is larger than would typically expected for a long GRB host correlation, however, we note there are significant errors on the individual physical separations.

Similarly to the full sample of our cross-matched sources, this final sample is dominated by ULIRGs. Two sources have radio spectral indices and IR colours consistent with star-forming galaxies, although we note that our sources' spectral indices are generally poorly constrained. We find a mean SFR of $59.2 \pm 36.1$ \sfrunits, significantly greater than that exhibited by the general population population of field galaxies and the CCSNe host sample.

\section{Conclusions}\label{sec:conc}

We have presented the results of a search for GRB host galaxies using the recent LoTSS DR2 catalogues. Using the density of sources in LoTSS DR2 to evaluate \pchance, we identified 18 sources matched to 17 GRBs. This process indicated crossmatches at relatively large angular and normalised separations. This was likely to be partly an effect of the differences between optical and radio galaxy morphology but also implied some inaccurate crossmatches. We further evaluated the properties of the sources using both LoTSS data and that available in other catalogues.

We found that a majority of our sources are consistent with ULIRG classifications, while a small minority are more likely to be AGN and unrelated to the matched GRBs. Our comparison CCSN host sample was also dominated by ULIRGs. We evaluated the star formation of our long GRB host sample, finding that they exhibited SFRs significantly higher (mean $66.8 \pm 38.2$ \sfrunits) than that of both field galaxies (mean $18.2 \pm 33.1$ \sfrunits) and the CCSN host sample (mean $8.5 \pm 23.3$ \sfrunits).

Based on both the results of our crossmatching process and the properties of the sources themselves, we have identified a final sample of eight crossmatches that are likely to be accurate host galaxy-GRB pairings. This sample consists entirely of long GRBs and is again dominated by ULIRGs exhibiting enhanced star formation (mean $59.2 \pm 36.1$ \sfrunits). 

Future observations by LOFAR will expand this sample over the coming years while also reaching to even lower frequencies with the low-band antennas. This will enable a more complete picture of GRB hosts in this regime to be developed and determine how their properties influence and drive the formation of GRB progenitors.

\section*{Acknowledgements}

We thank Emily Eyles-Ferris, Klaas Wiersema and Beatriz Mingo for invaluable discussion and insight.

We also thank the anonymous referee for their useful comments and improvements.

RAJEF acknowledges funding from the UK Space Agency and the European Union’s Horizon 2020 Programme under the AHEAD2020 project (grant agreement number 871158). RLCS acknowledges STFC support.

This work made use of data supplied by the UK \textit{Swift} Science Data Centre at the University of Leicester; the VizieR catalogue access tool, CDS, Strasbourg, France (DOI : 10.26093/cds/vizier); the Wide-field Infrared Survey Explorer, which is a joint project of the University of California, Los Angeles, and the Jet Propulsion Laboratory/California Institute of Technology, funded by the National Aeronautics and Space Administration.

Analysis of GRB 050509B is based in part on observations collected at the European Southern Observatory, Paranal, Chile (ESO program 075.D-0261, PI J. Hjorth).

LOFAR is the Low Frequency Array designed and constructed by ASTRON. It has observing, data processing, and data storage facilities in several countries, which are owned by various parties (each with their own funding sources), and which are collectively operated by the ILT foundation under a joint scientific policy. The ILT resources have benefited from the following recent major funding sources: CNRS-INSU, Observatoire de Paris and Universit\'{e} d'Orl\'{e}ans, France; BMBF, MIWF-NRW, MPG, Germany; Science Foundation Ireland (SFI), Department of Business, Enterprise and Innovation (DBEI), Ireland; NWO, The Netherlands; The Science and Technology Facilities Council, UK; Ministry of Science and Higher Education, Poland; The Istituto Nazionale di Astrofisica (INAF), Italy.

This research made use of the Dutch national e-infrastructure with support of the SURF Cooperative (e-infra 180169) and the LOFAR e-infra group. The J\"{u}lich LOFAR Long Term Archive and the German LOFAR network are both coordinated and operated by the J\"{u}lich Supercomputing Centre (JSC), and computing resources on the supercomputer JUWELS at JSC were provided by the Gauss Centre for Supercomputing e.V. (grant CHTB00) through the John von Neumann Institute for Computing (NIC).

This research made use of the University of Hertfordshire high-performance computing facility and the LOFAR-UK computing facility located at the University of Hertfordshire and supported by STFC [ST/P000096/1], and of the Italian LOFAR IT computing infrastructure supported and operated by INAF, and by the Physics Department of Turin university (under an agreement with Consorzio Interuniversitario per la Fisica Spaziale) at the C3S Supercomputing Centre, Italy.

\section*{Data Availability}

LOFAR LoTSS data are available at \url{https://lofar-surveys.org/}. 
\textit{Swift}-XRT data are available at \url{https://www.swift.ac.uk}. 
ESO VLT data are available at \url{https://archive.eso.org/}. 
The \textit{WISE}, NVSS, WENSS, FIRST, VLASS and other catalogued data used in this work are available via the VizieR library at \url{https://vizier.cds.unistra.fr}, DOI : 10.26093/cds/vizier.



\bibliographystyle{mnras}
\bibliography{bibs/grbs,bibs/LoTSS}



\appendix
\section{Source images}

\begin{figure*}
    \centering
    \includegraphics[width=\columnwidth]{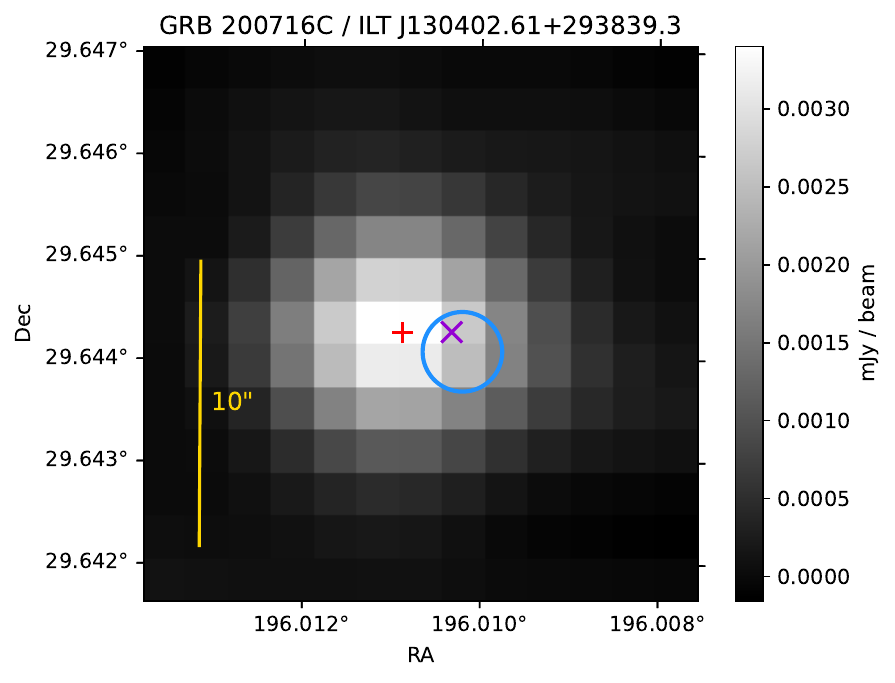}
    \includegraphics[width=\columnwidth]{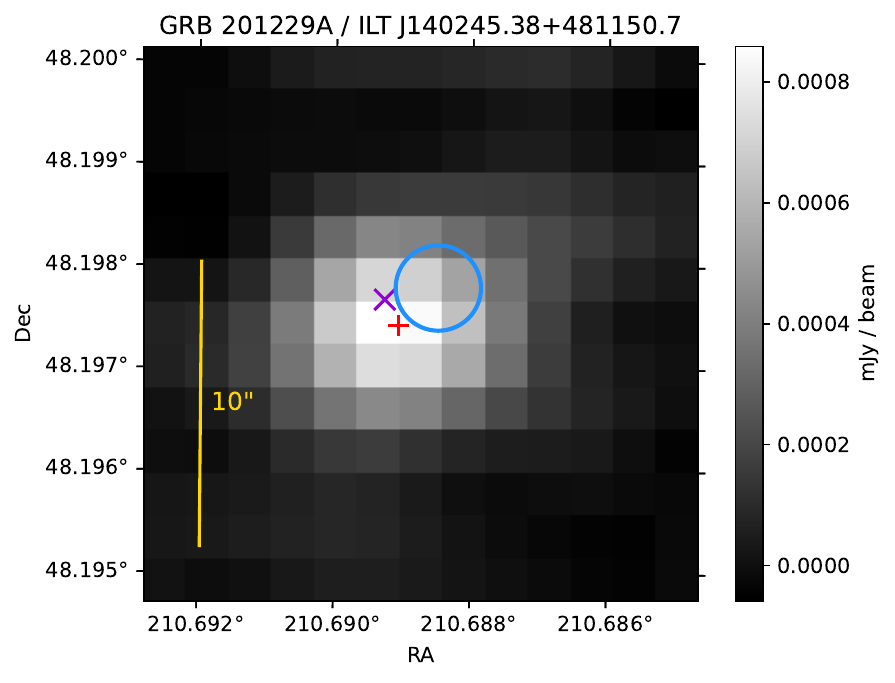}
    \includegraphics[width=\columnwidth]{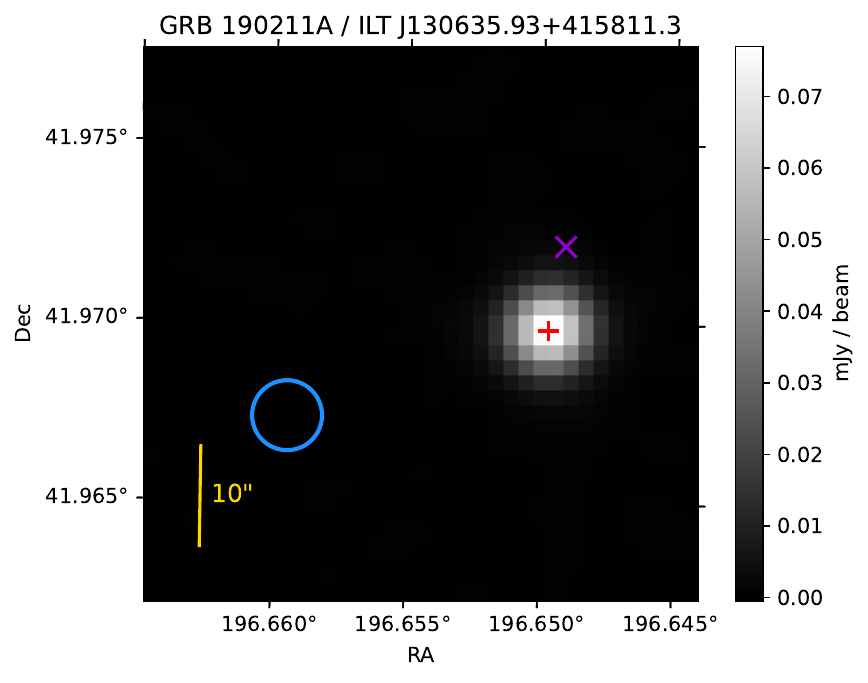}
    \includegraphics[width=\columnwidth]{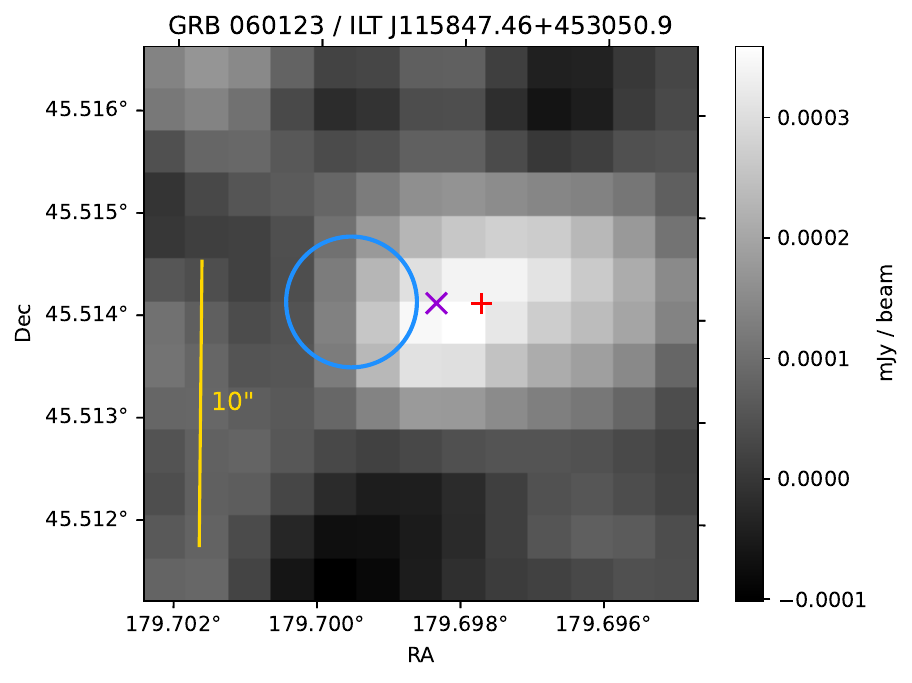}
    \includegraphics[width=\columnwidth]{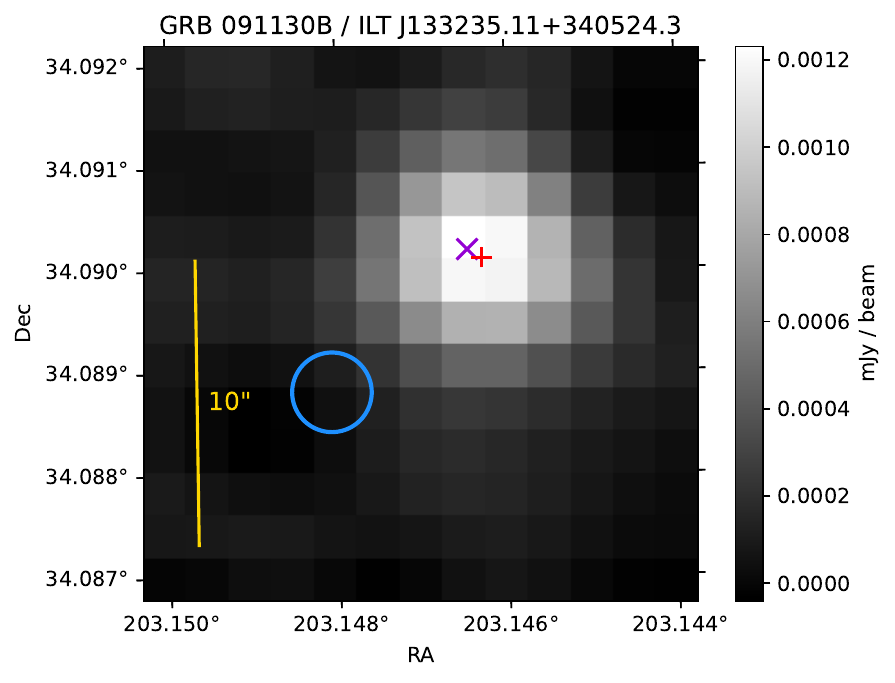}
    \includegraphics[width=\columnwidth]{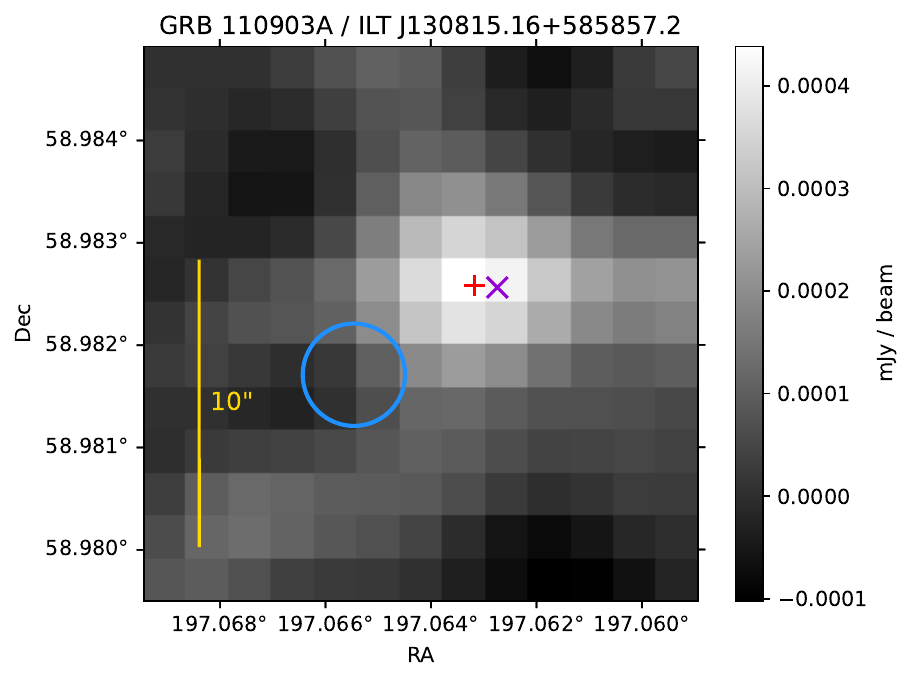}
    \caption{The LoTSS image mosaics for the matches to our GRB sample. All images show the full LoTSS band centred at 144 MHz and are ordered by GRB class and \pchance. The \textit{Swift}-XRT error region for each GRB is indicated with the blue circle, LoTSS matches are indicated with red $+$s and the AllWISE counterpart location to the LoTSS sources are indicated with purple $\times$s. Note the varying angular and flux density scales for each image. The LoTSS crossmatch is given in the title of each panel. For GRB 050509B, each crossmatch is labelled and for GRB 081025, each component of ILT J162131.13+602837.1 is labelled.}
    \label{fig:mosaics}
\end{figure*}

\begin{figure*}
    \centering
    \includegraphics[width=\columnwidth]{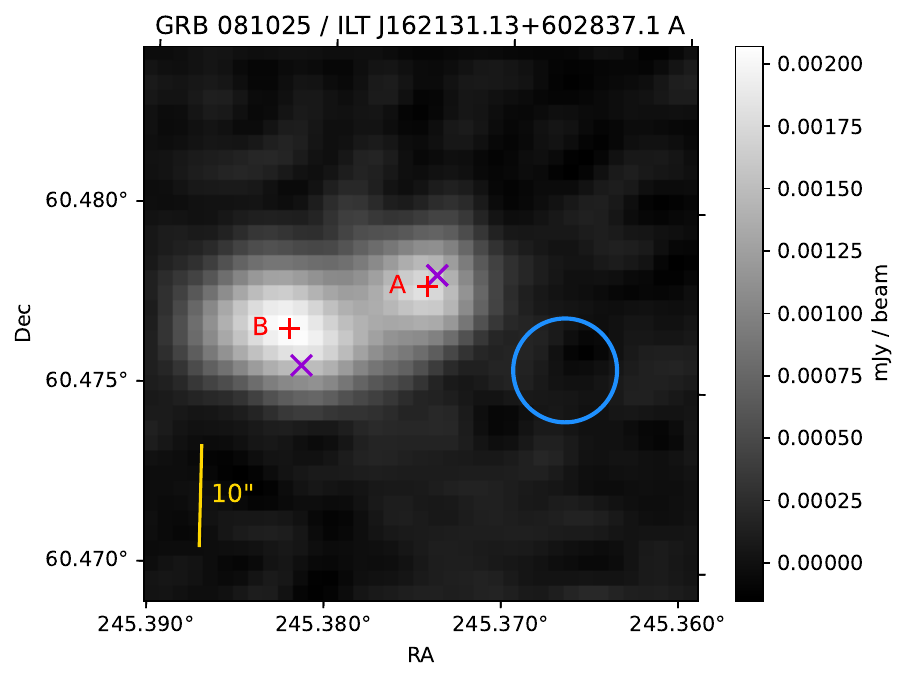}
    \includegraphics[width=\columnwidth]{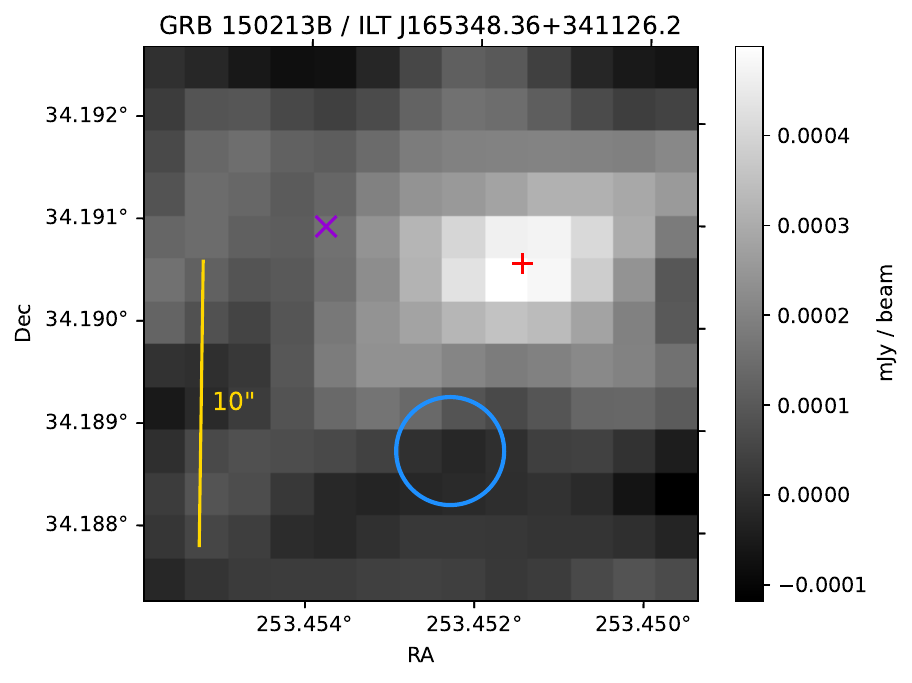}
    \includegraphics[width=\columnwidth]{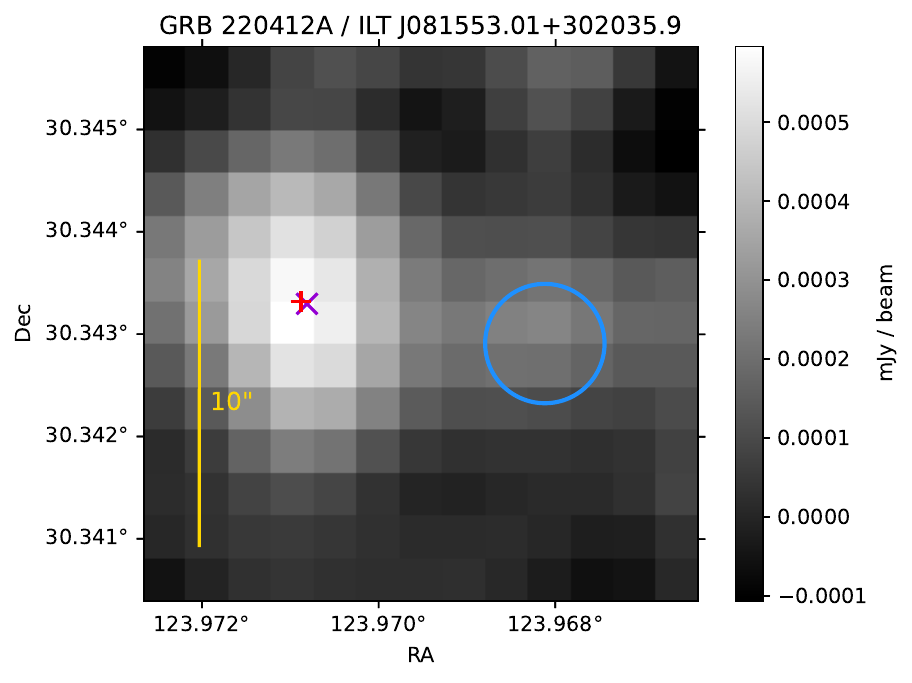}
    \includegraphics[width=\columnwidth]{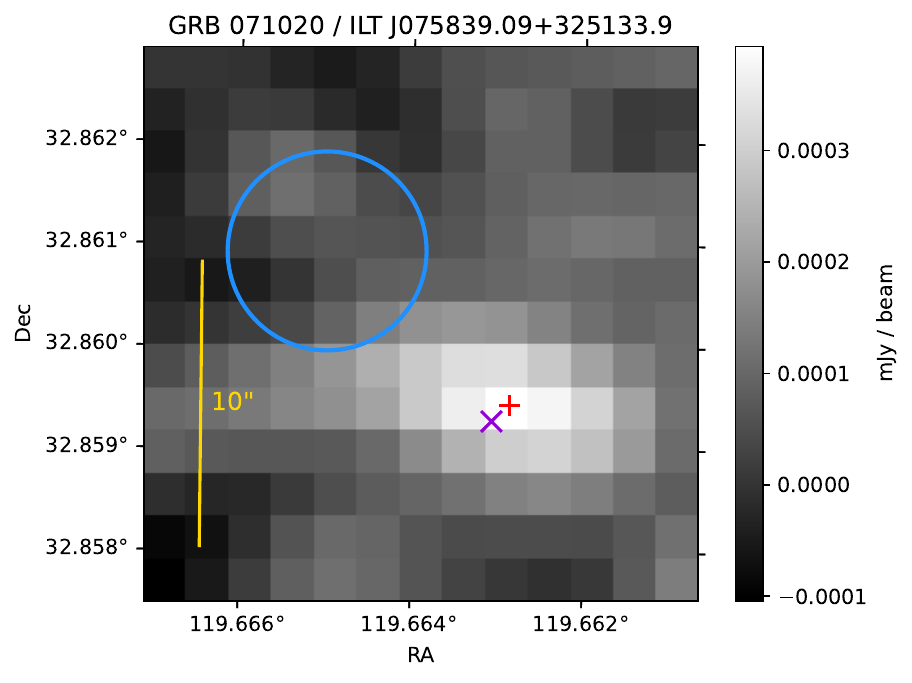}
    \includegraphics[width=\columnwidth]{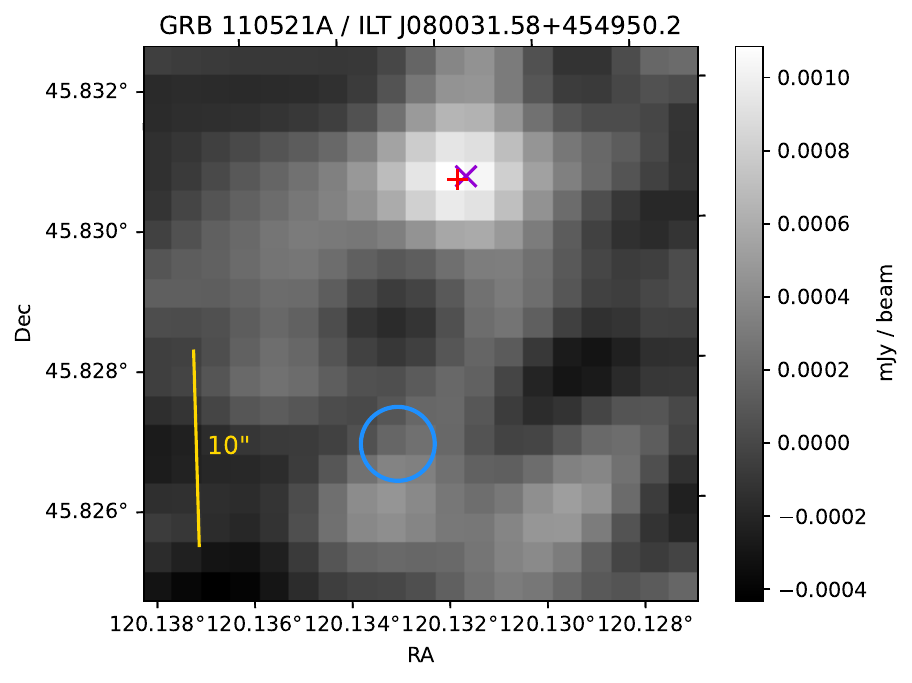}
    \includegraphics[width=\columnwidth]{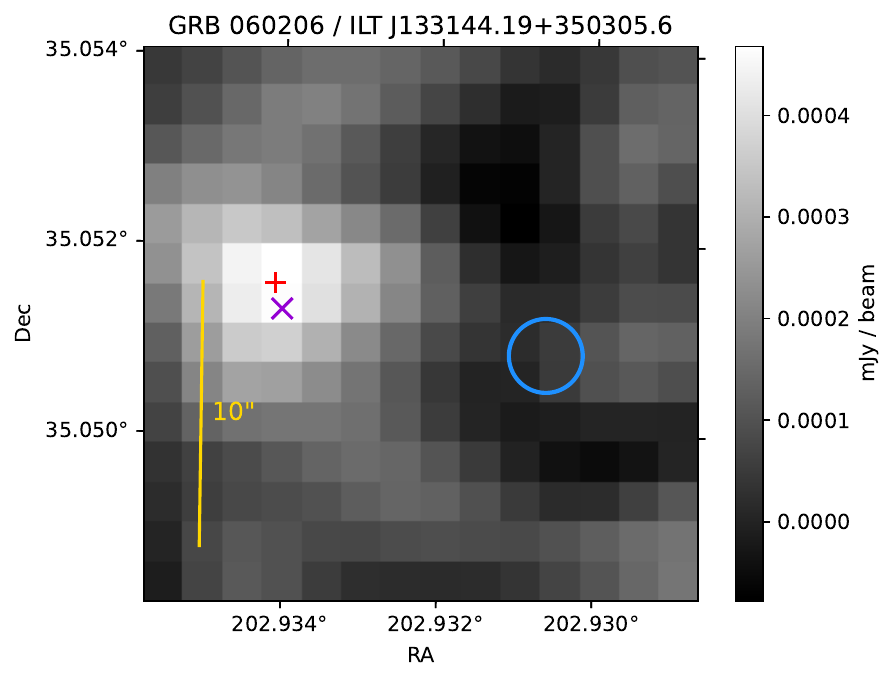}
    \contcaption{}
\end{figure*}

\begin{figure*}
    \centering
    \includegraphics[width=\columnwidth]{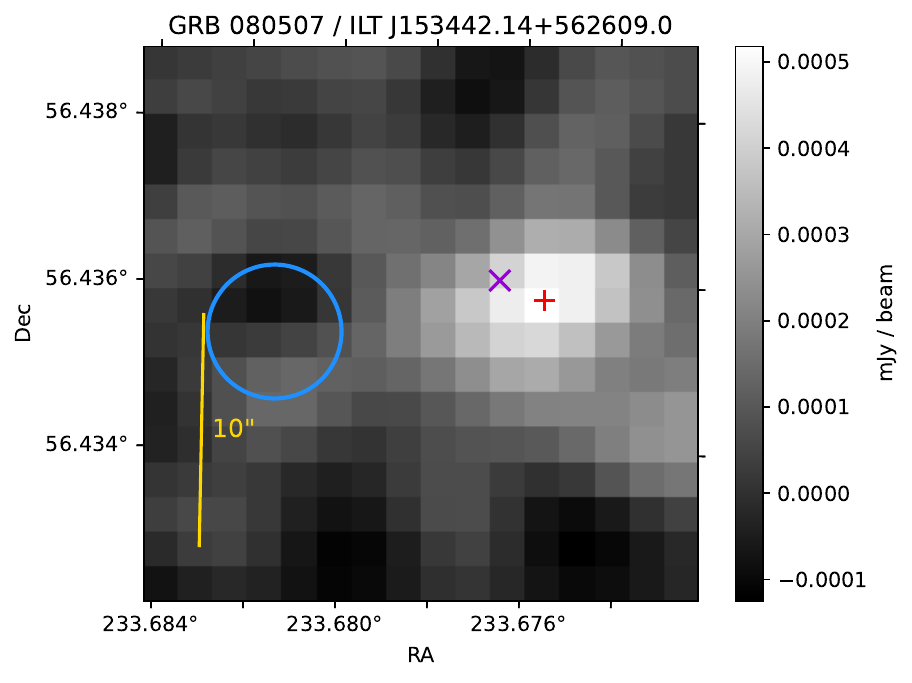}
    \includegraphics[width=\columnwidth]{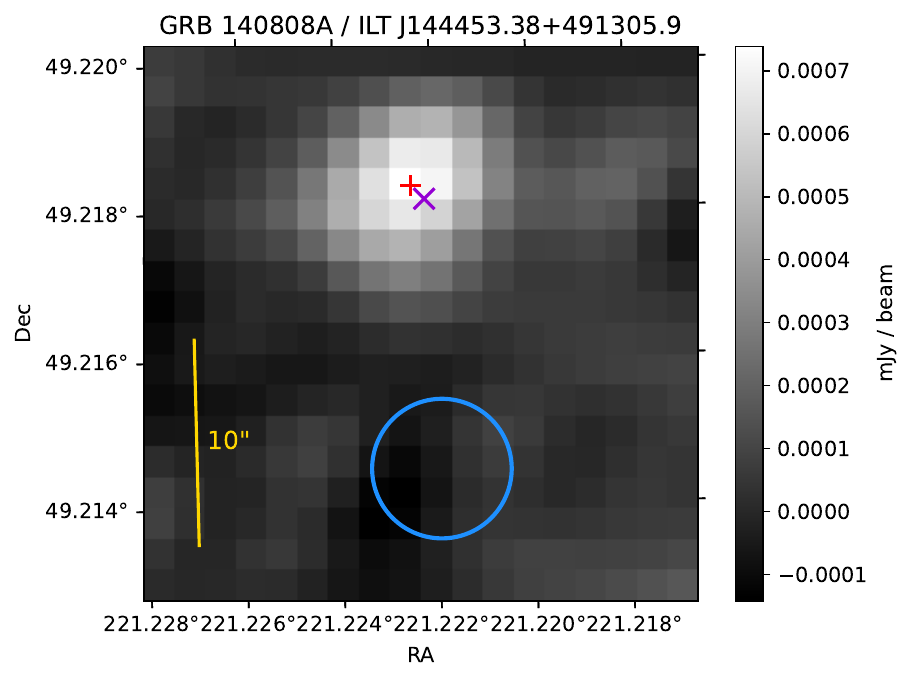}
    \includegraphics[width=\columnwidth]{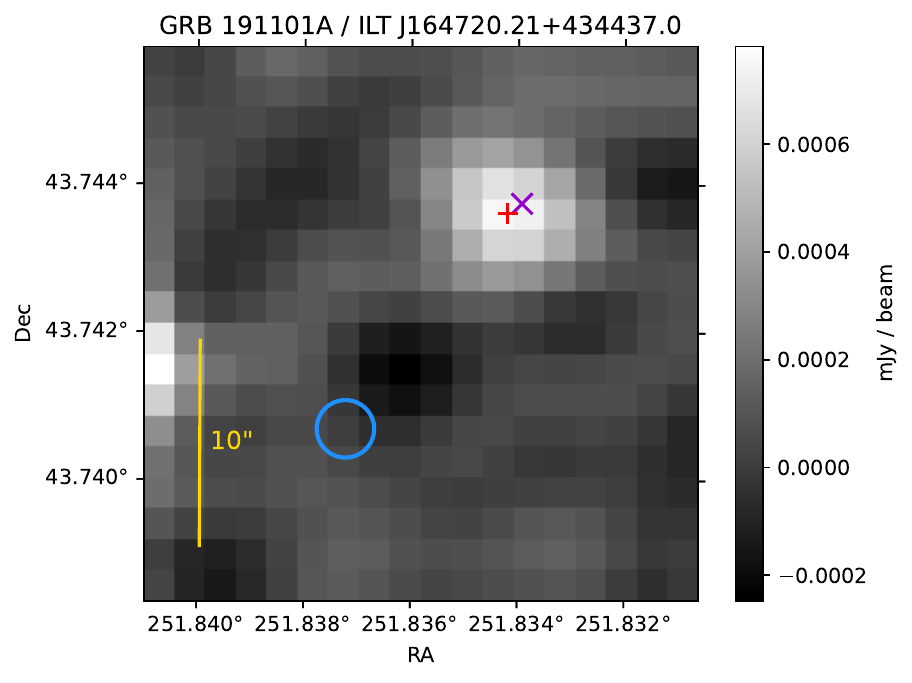}
    \includegraphics[width=\columnwidth]{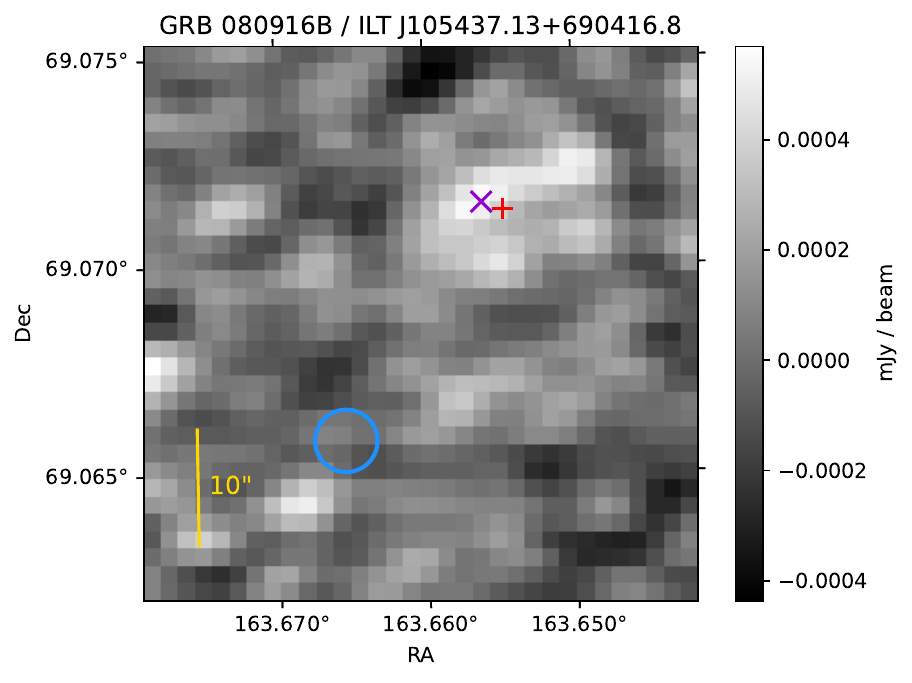}
    \includegraphics[width=\columnwidth]{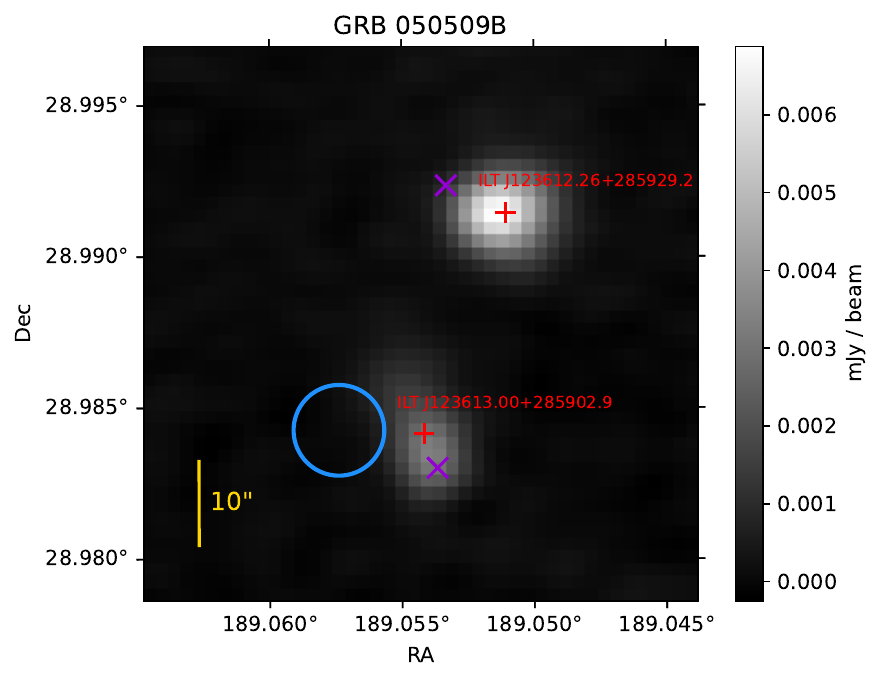}
    \contcaption{}
\end{figure*}

\begin{figure}
    \centering
    \includegraphics[width=\columnwidth]{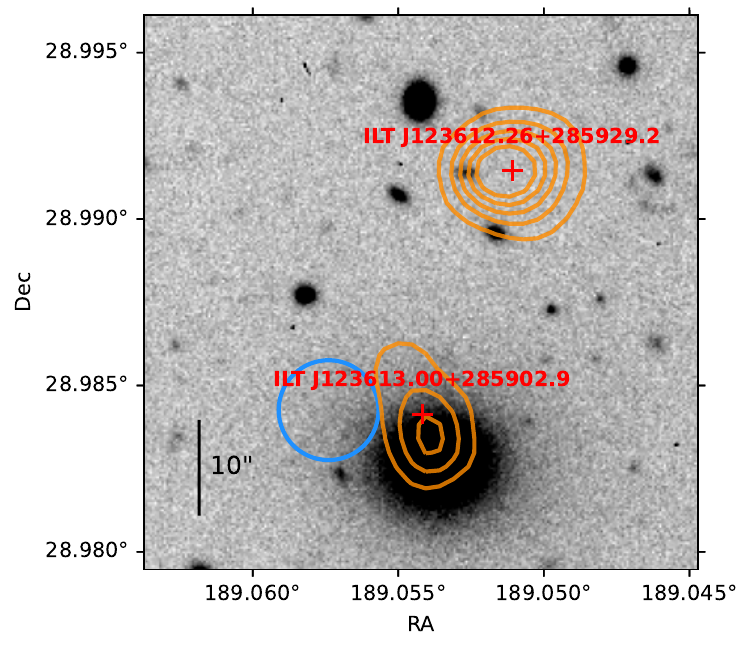}
    \caption{The VLT image of the region around GRB 050509B and its matched LoTSS sources \citep{Hjorth05a,Hjorth05,Gehrels05}. The \textit{Swift}-XRT error region is indicated with the blue circle, the LoTSS matches are indicated with the red $+$s and the orange contours indicate 0.001, 0.002, 0.003 and 0.004 mJy/beam in the LoTSS mosaic. ILT J123613.00+285902.9 is clearly coincident with the large elliptical galaxy assumed to be GRB 050509B's host.}
    \label{fig:050509b}
\end{figure}

\begin{figure}
    \centering
    \includegraphics[width=\columnwidth]{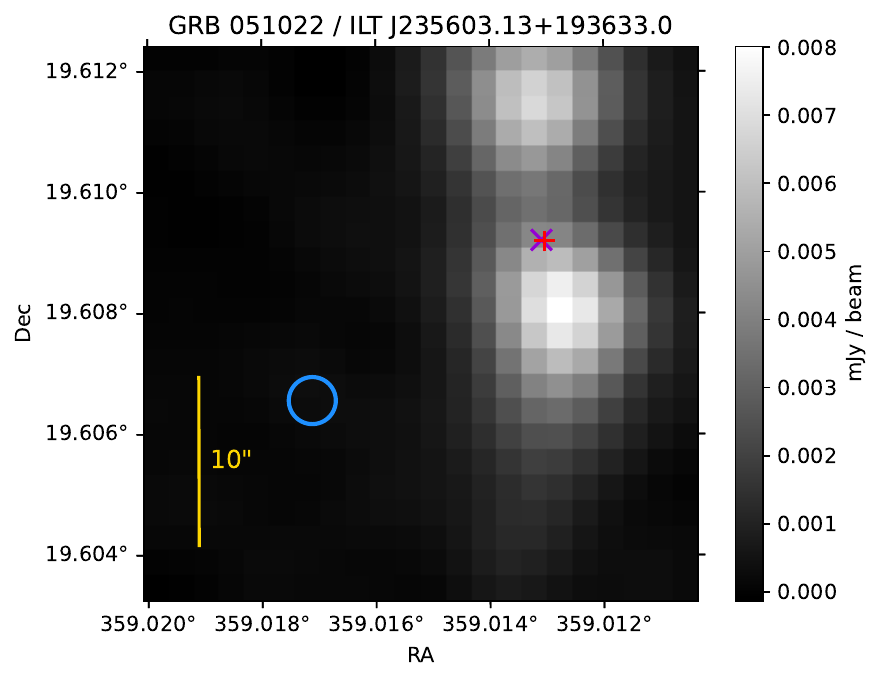}
    \caption{The LoTSS image mosaic around GRB 051022. The mosaic is the full LoTSS band centred at 144 MHz. The \textit{Swift}-XRT error region is indicated with the blue circle, the LoTSS match is indicated with the red $+$ and the AllWISE counterpart location is indicated with the purple $\times$.}
    \label{fig:051022}
\end{figure}

\section{Spectral energy distributions}
\begin{figure*}
    \centering
    \includegraphics[width=\columnwidth]{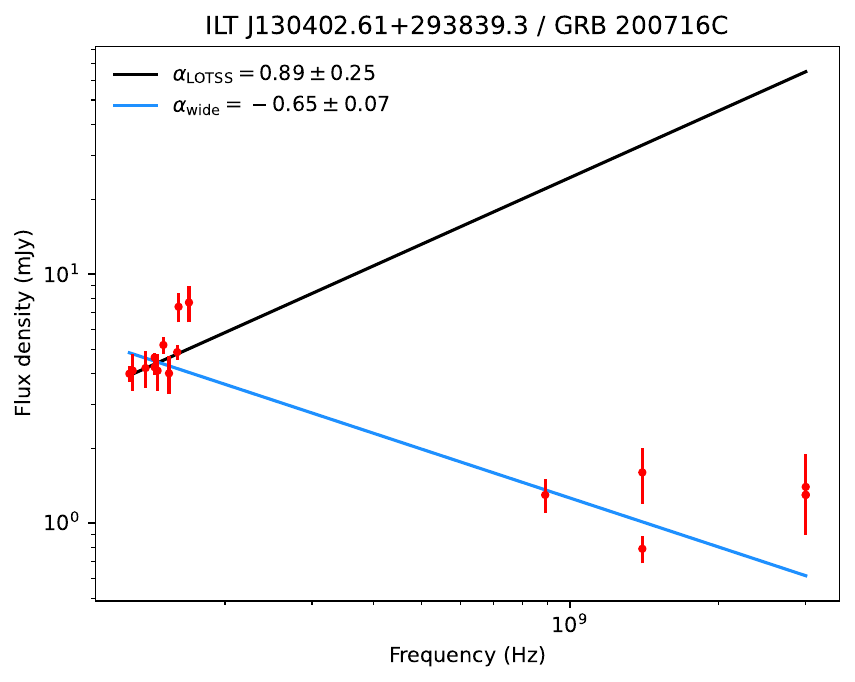}
    \includegraphics[width=\columnwidth]{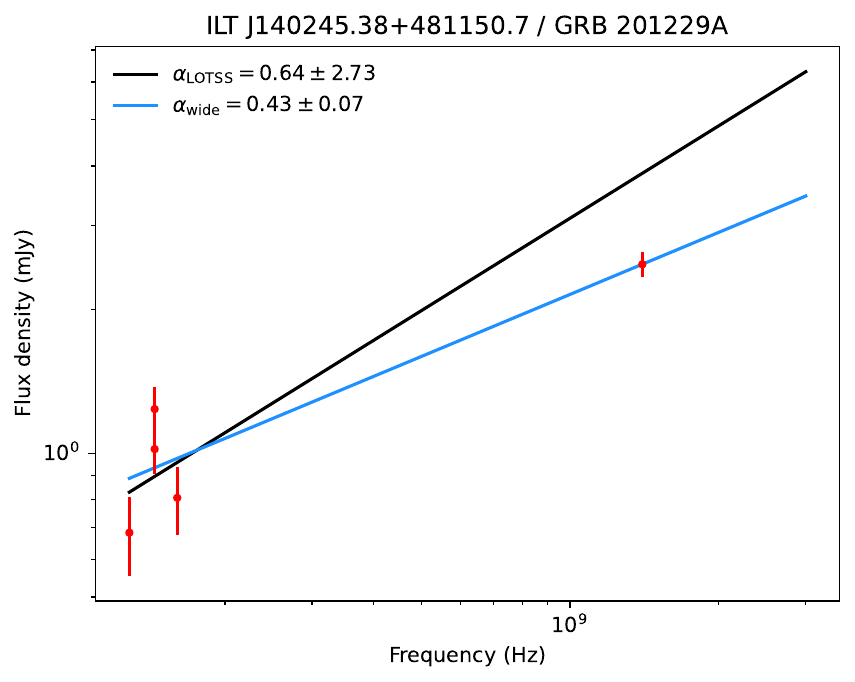}
    \includegraphics[width=\columnwidth]{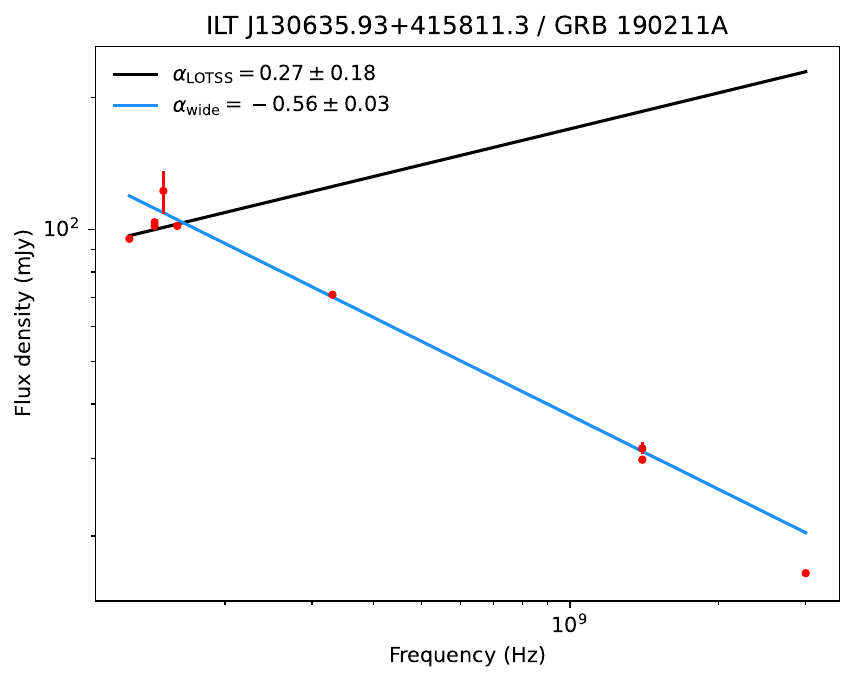}
    \includegraphics[width=\columnwidth]{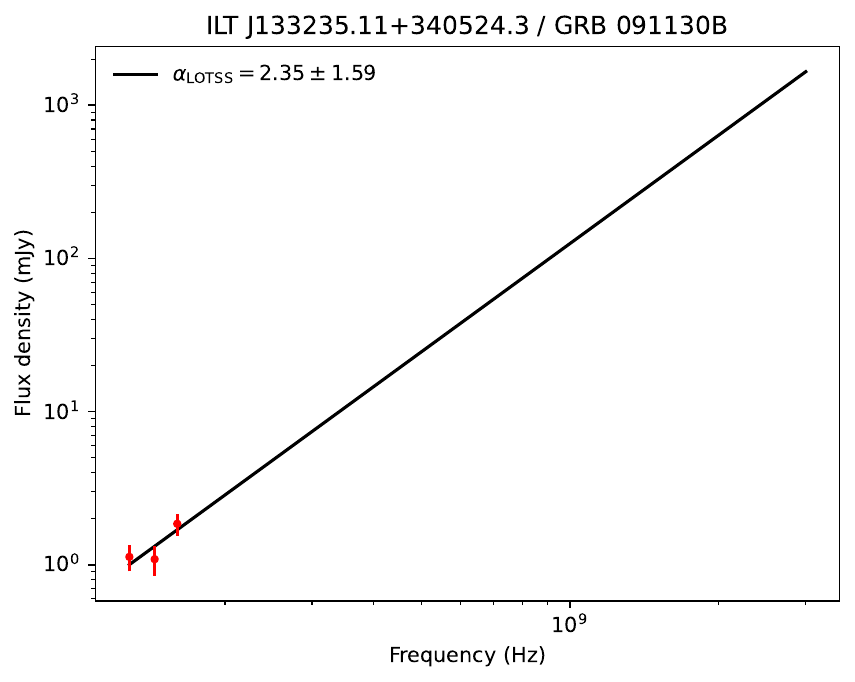}
    \includegraphics[width=\columnwidth]{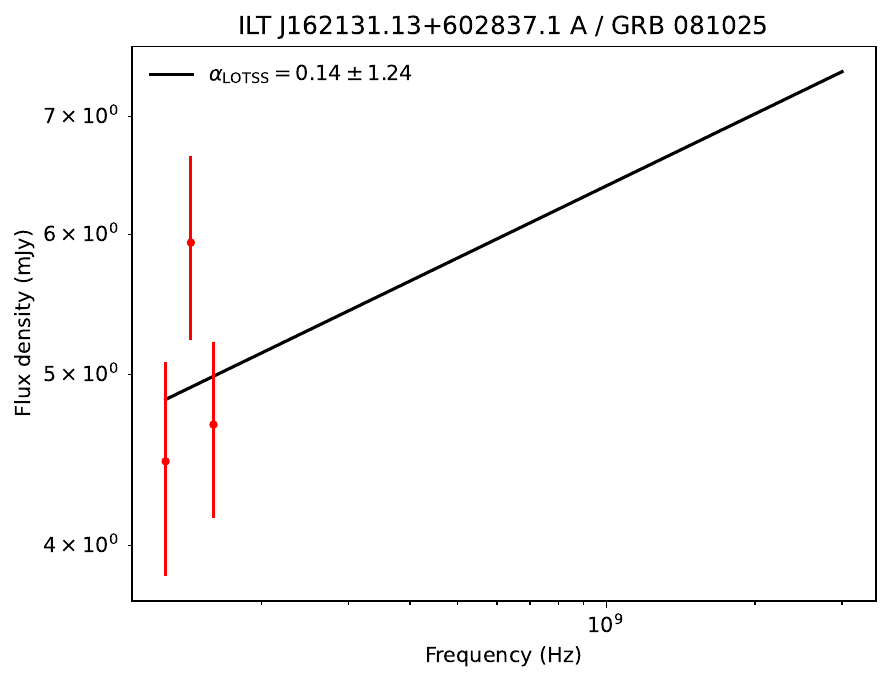}
    \includegraphics[width=\columnwidth]{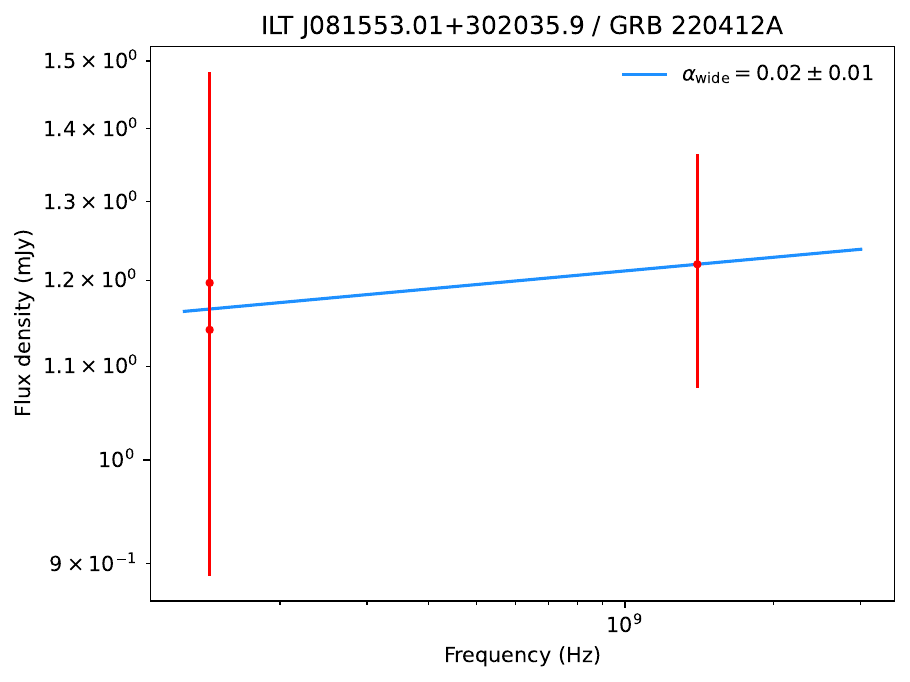}
    \caption{The SEDs of our matched sources over 128 MHz to 3 GHz ordered by GRB class and \pchance. Power laws fitted with only the LoTSS in-band flux densities are shown in black and those fitted with additional data from other catalogues are shown in blue.}
    \label{fig:seds}
\end{figure*}

\begin{figure*}
    \centering
    \includegraphics[width=\columnwidth]{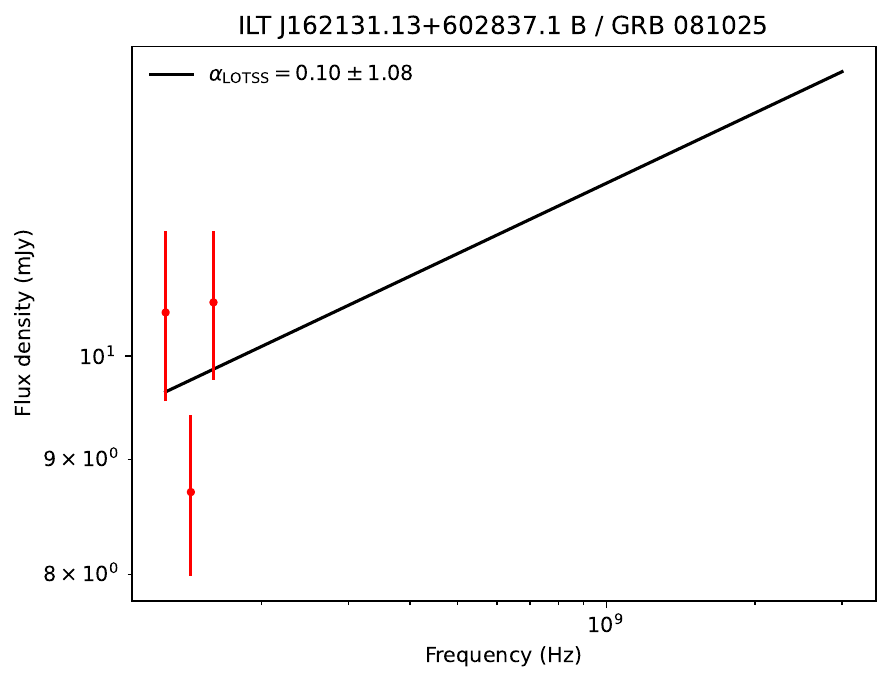}
    \includegraphics[width=\columnwidth]{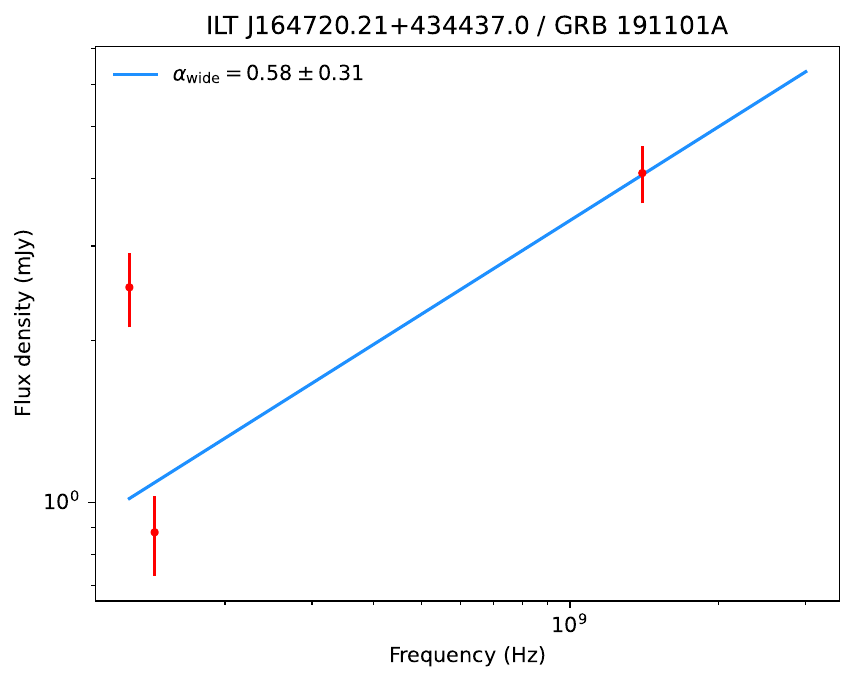}
    \includegraphics[width=\columnwidth]{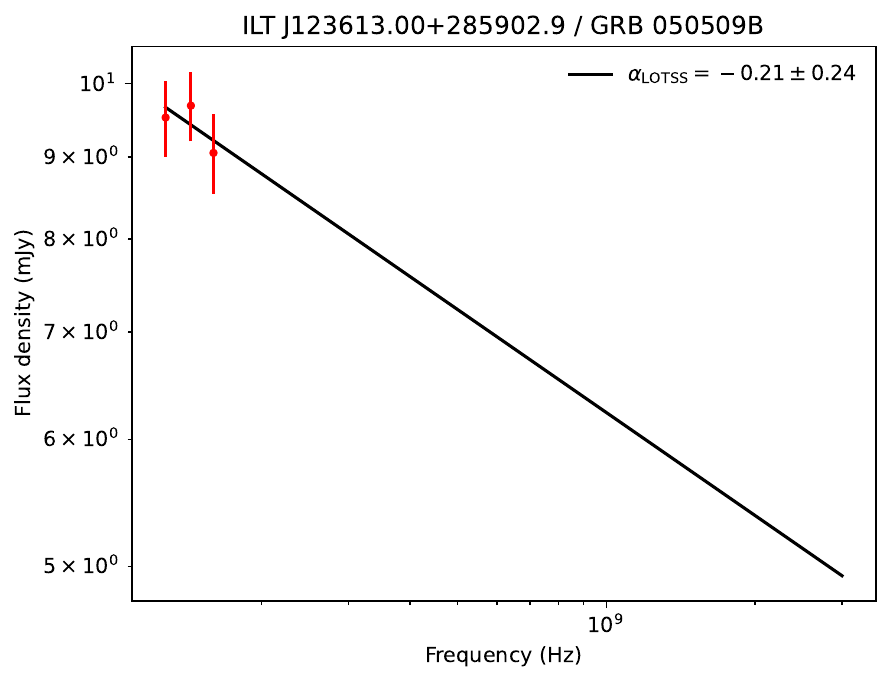}
    \includegraphics[width=\columnwidth]{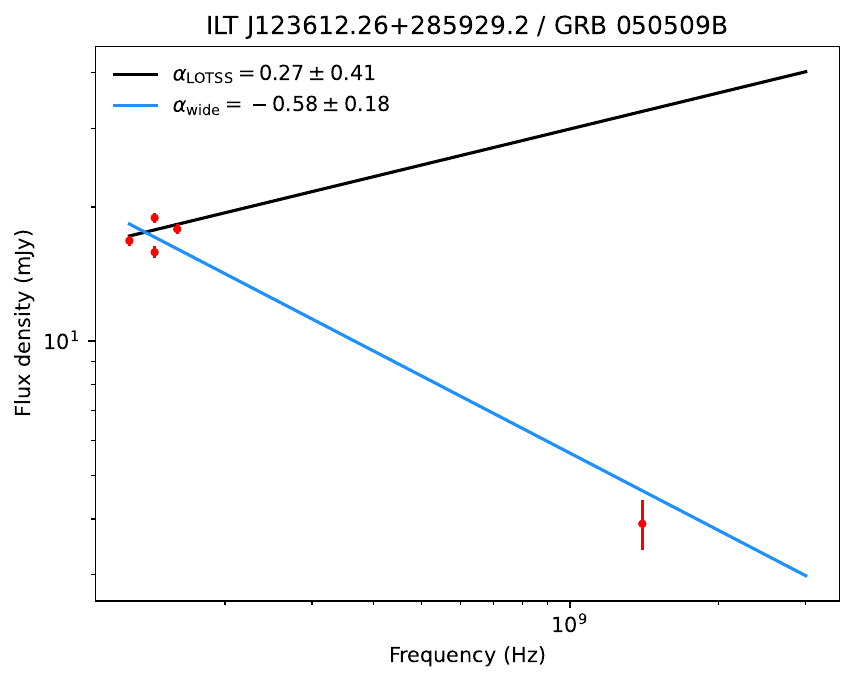}
    \contcaption{}
\end{figure*}


\bsp	
\label{lastpage}
\end{document}